\documentstyle[aps,prb,epsf]{revtex}
\draft
\newcommand{\grad}[0]{\mbox{\boldmath $\nabla$}}
\newcommand{\bx}[0]{{\bf x}}
\newcommand{\br}[0]{{\bf r}}
\newcommand{\bR}[0]{{\bf R}}
\newcommand{\bF}[0]{{\bf F}}
\newcommand{\bbf}[0]{{\bf f}}
\newcommand{\bn}[0]{{\bf n}}

\newcommand{\bK}[0]{{\bf K}}
\newcommand{\brho}[0]{\bar{\rho}}
\newcommand{\bj}[0]{{\bf j}}
\newcommand{\bu}[0]{{\bf u}}
\newcommand{\bw}[0]{{\bf w}}
\newcommand{\ba}[0]{{\bf a}}

\newcommand{\tf}[0]{\widetilde{{\bf f}}}
\newcommand{\fpin}[0]{{\bf f}_{\rm pin}}
\newcommand{\trho}[0]{\widetilde{\rho}}
\newcommand{\tj}[0]{\widetilde{{\bf j}}}

\newcommand{\mG}[0]{{\cal G}}
\newcommand{\cR}[0]{{\cal R}}

\begin{document}
\twocolumn[\hsize\textwidth\columnwidth\hsize\csname @twocolumnfalse\endcsname

\title{Static and Dynamic Properties of\\
Inhomogeneous Elastic Media on  Disordered Substrate }

\author{Dinko Cule and Terence Hwa}

\address{Physics Department, University of California at San Diego,
         La Jolla, CA 92093-0319}
\date{\today}
\maketitle

\begin{abstract}
The pinning of an inhomogeneous elastic medium by a disordered substrate 
is studied analytically and numerically. The static and dynamic properties of 
a $D$-dimensional system are shown to be equivalent to those of the 
well known problem of a $D$-dimensional random manifold embedded 
in $(D+D)$-dimensions. The analogy  is found to be very robust, 
applicable to a wide range of elastic media,  including those which are 
amorphous or nearly-periodic, with local or nonlocal elasticity. 
 Also demonstrated explicitly is the equivalence between 
the dynamic depinning transition obtained at a constant driving force, 
and the self-organized, near-critical behavior 
obtained by a (small) constant velocity drive.
\end{abstract}

]

\section{Introduction}

\label{S1}

The pinning of elastic continuum in random potential has been a subject of
numerous studies in the past decade~\cite{ke1}. 
It is related to various phenomena of  technological importance, 
while being also of fundamental interest to the statistical mechanics of
disordered systems. Considerable efforts have been devoted to understanding
how quenched impurities influence the transport of charge-density wave (CDW)~%
\cite{cdw,cdw1,nf}, pinning of flux lines in type-II superconductors~\cite{blatter},
roughness of crystalline surfaces~\cite{td,ts}, propagation of invasion 
fronts~\cite{front},
etc. Many aspects of these systems can be described either by the model of
randomly-pinned CDW or by the model of randomly-pinned directed manifold
(``random manifold'' for brevity). The static properties of the
low-temperature glass phases of these systems have been studied by a variety
of analytical methods, including renormalization-group (RG) 
analysis~\cite{co,psrg}, replica variational method~\cite{rft1,gld},
and functional RG~\cite{frg,kor,gld}. Combined with the exact ground
state structures obtained numerically via efficient (polynomial) 
algorithms~\cite{tm,dp,aam,mcmf},
these systems are perhaps the best characterized glassy system to date. Much
progress have also been made in understanding the nonequilibrium driven
dynamics of these systems~\cite{ke1}. 
In the extreme nonequilibrium limit where thermal
fluctuations can be neglected, it is known that the driving force exceeding
a critical value is necessary to depin the system. A continuous dynamic
phase transition occurs at the depinning threshold, where the dynamics
exhibits complex stick-slip motion with ``avalanches'' of all sizes. Such
complex dynamics results from the intricate balance of elasticity and random
pinning forces near the onset of motion. They have been characterized in
great details by a combination of analytic~\cite{nf,natt,nf2,ke2}, 
numeric~\cite{pbl,mf,myers,robbins,string,lt,stanley,cule}, 
and experimental~\cite{rubio,hor,wong,bu,family} methods.

Another class of pinning phenomena which have attracted much attention 
is the tribology of sliding elastic
bodies interacting via a contact surface. This is exemplified by the
Burridge-Knopoff model~\cite{bk,taka,cl,naka} 
describing the dynamics along an individual
earthquake fault. Other examples include boundary layer lubrication~\cite{lub},
and stick-slip motion~\cite{feder,gollub} of elastic continuum 
over sticky substrates. Complex spatio-temporal dynamics are found to occur
in these systems also, when they are driven slowly. 
Previous studies~\cite{cls} of this class of systems have focused 
on the role of chaos generated by nonlinear dynamics, e.g., 
a velocity-weakening friction.
Complexity in these systems are not as well understood theoretically.

Another approach to studying the spatio-temporal complexity generated
in tribology-like problems is by ``sandpile''-like automaton 
models, as pioneered by Bak and collaborators~\cite{bt,cbo}.
The connection between sandpile-like models and dynamic critical
phenomena has been proposed early on by Tang and Bak~\cite{tb}. 
(In fact, the sandpile model~\cite{btw} itself was motivated from
the study of  the randomly-pinned CDW~\cite{twbcl}.
Equivalence between many aspects of the
sandpile model and those of the CDW at the depinning threshold has since been 
verified numerically by Narayan and Middleton~\cite{nm}.)
Analogy between the geometrical structures of individual earthquake fault zones
and those of the {\em equilibrium} random manifold has also been 
explored~\cite{sornette}.
In a previous letter~\cite{prl}, 
we demonstrated analytically and numerically the
equivalence between a certain class of disorder-dominated tribology problem
and the depinning dynamics of the random manifold. The same approach has
been used  by Fisher {\it et al}~\cite{fisher} as a starting point in analyzing
earthquake fault dynamics. Connection between 
a related interface depinning problem and certain earthquake models 
have also been discussed by Paczuski and Boettcher~\cite{pb}. 
  In this paper, we examine in detail the relation 
between the tribology-like systems and the randomly-pinned CDW/manifolds. 
Our major result, that 
{\em both the static and dynamic properties of a $D$-dimensional
inhomogeneous elastic body embedded in a $D$-dimensional random medium
are equivalent to those of a homogeneous $D$-dimensional directed
manifold embedded in $(D+D)$-dimensional random medium}, is found to
be very robust, applicable to a wide range of elastic media, including those
which are amorphous, nearly-periodic, with local or nonlocal elasticity.
These results are relevant to a number of apparently unrelated problems,
including the enhanced pinning of entangled flux lines, the nonequilibrium
freezing of moving vortex array~\cite{hmv}, 
the reptation of heteropolymers~\cite{reptate}, and
alignment of DNA sequences~\cite{align}. 
We also demonstrate explicitly the
connection between the dynamic critical phenomena obtained from a constant
driving force at the depinning threshold, and the self-organized,
nearly-critical behavior obtained by a (small) constant 
velocity drive. The latter is an example of ``extremal dynamics'' 
by which many self-organized critical phenomena arise~\cite{pmb}.

This paper is organized as follows. We first introduce in Sec.~II a discrete 
$D$-dimensional model of inhomogeneous elastic system pinned in a disordered
medium, and derive the appropriate continuum Hamiltonian. 
The statics of the continuum system is examined analytically
in Sec.~III.A. We show that while a perfectly periodic system is
equivalent to a randomly-pinned CDW system, any quenched-in inhomogeneity
changes the universality class to that of the random manifold. This includes
somewhat surprisingly even nearly-periodic systems with a small
concentration of quenched-in interstitials and vacancies, or with only
quenched-in phonon modes. This finding is
demonstrated numerically by detailed analysis of the model in $D=1$. All
universal quantities examined, including amplitude ratios, are found to be
indistinguishable from those of the one-dimensional random manifold, i.e.,
the randomly pinned directed path in $1+1$ dimensions. The driven dynamics
of the elastic system (with a constant force) is described next in Sec.~IV.
The random manifold analogy is extended to include critical depinning
dynamics, and demonstrated explicitly for the case $D=1$. In Sec.~V, we study
the effect of nonlocal elasticity mediated by the elastic body in the bulk
not in contact with the disordered substrate. The dynamics is shown
to be analogous to the appropriate random manifold problem with nonlocal
elasticity. Finally, we compare the dynamics obtained
 with constant-force drive and
that with constant-velocity drive. We find the two to be nearly equivalent
close to the depinning threshold where the average motion is slow. In order for
this manuscript to be self-contained, we provide brief reviews of the basic
properties concerning the statics and dynamics of the CDW/random manifold
systems within the text. More detailed explanations of the coarse-graining
procedure and a review of bulk-mediated elasticity are provided in 
Appendix A and B.

\section{The Discrete and Continuum Models}

\begin{figure}
\narrowtext
\epsfxsize=2.0truein
\centerline{\epsffile{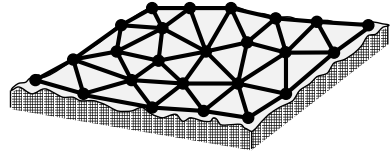}}
\vspace{\baselineskip}
\caption{
A  randomly-tethered elastic membrane in contact with
a disordered substrate. 
%The configuration is
%specified by the set of position vectors $\{\br_n\}$ for the beads 
%labeled by $\{n\}$.
}
\label{F01}
\end{figure}

In this section, we considered a $D$-dimensional {\em inhomogeneous} elastic
medium, e.g., an amorphous solid, which is completely immersed in a 
$D$-dimensional disordered substrate. An example for the $D=2$ case is
a sheet of latex membrane in contact with glass~\cite{gollub}, or a 
randomly-polymerized membrane adsorbed on a substrate (see Fig.~\ref{F01}).
This situation also
arises in the pinning of a rigid vortex array (with quenched-in defects) in
a thin-film superconductor. Moreover, $D=3$ may describe the pinning of an
entangled vortex line network in a bulk superconductor (see below), 
and $D=1$ is relevant to the reptation of a heteropolymer 
in a disordered gel matrix~\cite{reptate}.

We use an irregular array of beads tethered by harmonic springs to model the
amorphous solid. Let the average inter-bead distance
be $a$ and the {\em equilibrium} position of a bead labeled by $n$ be given by 
${\bf R}_n\in \Re ^D$. The neighboring beads are connected by harmonic
springs with appropriately chosen spring lengths (of the order $a$) such
that the configuration $\left\{ {\bf R}_n\right\} $ is the {\em unfrustrated}
ground state of the tethered system in the absence of any external forces.

To describe the large scale properties of such an elastic medium
analytically, it will be useful to adopt a continuum description.
Let the density field of the {\em unperturbed} system be
\begin{equation}
\rho_0(\br) = \sum_{\{n\}}\delta ^D({\bf r}-{\bf R}_n).  \label{rho.0}
\end{equation}
The quenched-in density variation is $\delta \rho _0({\bf r%
})=\rho _0({\bf r})-\bar{\rho}$, where $\bar{\rho}\equiv a^{-D}$ is the
average density; it is characterized by the correlation
function 
\begin{equation}
\overline{\delta \rho _0({\bf r})\delta \rho _0({\bf r^{\prime }})}\equiv
C_0({\bf r}-{\bf r^{\prime }}),  \label{rho-rho}
\end{equation}
or the structure factor $S_0({\bf k})=\int d^D{\bf r}\,C_0({\bf r})e^{i{\bf k%
}{\bf r}}$. Here the overbar denotes average over the ensembles of quenched
bead positions $\{{\bf R}_n\}$. In the case of a single large system, the
overbar can be taken as the spatial average over smaller subsystems.

\begin{figure}
\narrowtext
\epsfxsize=1.6truein
\centerline{\epsffile{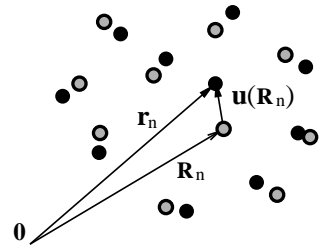}}
\vspace{\baselineskip}
\caption{
The {\em equilibrium} position (circle) of a bead $n$ is given $\bR_n$;
the actual position (black dot) $\br_n$ is given by the displacement 
vector $\bu(\bR_n)$. 
Note that the set $\{\bR_n\}$ is not necessarily ordered periodically.
}
\label{F01.b}
\end{figure}

Elastic distortion of the discrete system is described by a displacement
vector $\bu({\bf R}_n)$, which  denotes the displacement of the 
$n^{{\rm th}}$ bead from its equilibrium position ${\bf R}_n$; 
see Fig.~\ref{F01.b}.
 Then the
actual position of the bead is $\br_n={\bf R}_n+\bu(\bR_n)$, 
and the corresponding density field is 
\begin{eqnarray}
\rho(\br;\bu)&=&\sum_{\{n\}}\delta ^D\big({\bf r}-{\bf R}_n-
\bu(\bR_n)\big)\nonumber\\
&\approx& (1-\grad \cdot \bu) \sum_{\{n\}}\delta ^D\big({\bf r}-{\bf R}_n-
\bu(\br)\big),  \label{rho}
\end{eqnarray}
where we assumed that  variations in the displacement is small, i.e., $\grad
\cdot \bu \ll 1$. 
Comparing Eq.~(\ref{rho}) with Eq.~(\ref{rho.0}), we see that 
\begin{eqnarray}
\rho ({\bf r},{\bf u})&=&(1-\mbox{\boldmath$\nabla$}\cdot {\bf u})\cdot 
\rho_0({\bf r}-{\bf u})\nonumber\\
&\approx & \brho (1 - \grad \cdot \bu) + \delta\rho_0(\br-\bu)
\label{rho.2}
\end{eqnarray}
to leading order in $(\mbox{\boldmath$\nabla$}\cdot {\bf u})$. In the
continuum description, details of the elastic medium are contained
completely in the term $\delta\rho _0(\br)$, 
through the correlation function $C_0(\br)$.

In the absence of any external forces, the energy of the tethered system is
invariant upon a constant displacement ${\bf u}\to {\bf u}+const$. For a
statistically isotropic medium, the energy of small elastic distortion is
then simply given by the classical form~\cite{landau}
\begin{equation}
{\cal H}_0=\int d^D{\bf r}\left\{ \frac{c_{11}}2(\mbox{\boldmath $\nabla$}%
\cdot {\bf u})^2+\frac{c_{66}}2(\mbox{\boldmath $\nabla$}\times {\bf u}%
)^2\right\} ,  \label{H0}
\end{equation}
where $c_{11}$ and $c_{66}$ are the bulk and shear elastic moduli~\cite{N2}.

Consider now the situation where the elastic medium is in contact with a
disordered substrate, modeled by a Gaussian random potential $V({\bf r%
})$ with zero mean and a variance 
\begin{equation}
\left[ V({\bf r})V({\bf r^{\prime }})\right] =\Delta _V\,\delta ^D({\bf r}-%
{\bf r}^{\prime }).  \label{V-V}
\end{equation}
Here, $\left[ ...\right] $ denotes average over the ensemble of substrates.
The interaction of the medium with the substrate is described by
a pinning energy, 
\begin{eqnarray}
{\cal H}_{{\rm pin}} &=& \sum_{\{n\}} V(\br_n) 
= \int d^D{\bf r}\,\rho ({\bf r},{\bf u})V({\bf r})\nonumber \\
&=& \int d^D \br\,\Big\{ - \brho (\grad\cdot\bu) V(\br) 
+ W(\br,\bu(\br))\Big\},
\label{H.pin}
\end{eqnarray}
where 
\begin{equation}
W(\br,\bu)\equiv \delta \rho_0(\br-\bu) V(\br), \label{W}
\end{equation}
and the term $\brho V(\br)$ is neglected as it produces merely an overall
energy shift.
Note that $W(\br,\bu)$ depends explicitly on ${\bf u}$ where as  other
terms depend only on $\grad \, \bu$. Thus, 
only $W$ breaks the translational symmetry ${\bf u}\to {\bf u}+const$, 
and is responsible for providing the pinning phenomenon. The
interacting system, characterized by the effective Hamiltonian 
\begin{equation}
{\cal H}={\cal H}_0+\int d^D{\bf r}\, \Big\{ -\bar{\rho}\,
(\grad\cdot {\bf u}) V(\br) +W({\bf r},{\bf u})\Big\} ,  \label{H.eff}
\end{equation}
will be analyzed in detail in the next section~\cite{N3}.

\section{Thermodynamic Properties}

\subsection{Theoretical Considerations}

\subsubsection{The Periodic Medium}

If the tethered system does not contain any quenched-in defects, then the
intrinsic density variation is periodic, i.e., 
\begin{equation}
\delta \rho _0({\bf x})\sim \overline{\rho }\,\sum_i\cos \left( {\bf K}%
_i\cdot {\bf x}\right) ,  \label{drho.cdw}
\end{equation}
where ${\bf K}_i$'s are the reciprocal lattice vectors. Eq.~(\ref{H.eff}) in
this case reads 
\begin{eqnarray}
{\cal H}_{{\rm CDW}} &=&{\cal H}_0+\int d^D{\bf r}\Big\{ -\bar{\rho}\,
(\grad\cdot\bu)V(\br)   \nonumber \\
&&\quad  +\sum_i\brho \, V({\bf r})\,\cos \big[ {\bf K}_i\cdot 
{\bf (r-u(r)}\big] \Big\} .  \label{cdw}
\end{eqnarray}
The model (\ref{cdw}) is the $D$-component generalization of the
randomly-pinned charge-density wave (CDW) in $D$-dimensions~\cite{cn,cld}. 
The equilibrium properties of this class of systems have been well 
studied~\cite{co,psrg,gld}: It is known that
the disorder is irrelevant in $D=1$, where the system behaves like a
Gaussian chain. In $D>2$, the system (\ref{cdw}) is glassy at any finite
temperatures~\cite{psrg,gld}. 
The glass phase is described by two critical exponents: the
thermal exponent $\theta $ characterizing typical fluctuations of the free
energy landscape $\Delta {\cal F}\sim L^\theta $ for elastic distortion over
the length scales $L\gg a$, and the ``wandering'' exponent $\zeta $
characterizing fluctuations of the displacement field, $\Delta u\sim L^\zeta
,$ in the low free energy state(s). The results $\theta =D-2$ and $\zeta
=O(\log )$ is believed to be exact for $2<D\leq 4$. Right in $D=2$, the
situation is somewhat more complicated~\cite{co,rsb,vgnum}. 
The disorder is irrelevant for $T>T_g\propto 
\left( \brho\,\big( c_{11}^{-1}+c_{66}^{-1}\big)\right) ^{-1}$. 
The precise value of the glass temperature $T_g$ depends on
the microscopic model~\cite{cn,cld}. 
Below $T_g,$ the 2-dimensional system is in a
``marginal'' glass phase~\cite{co,cn,cld,hf} characterized 
by logarithmic anomalies in $\Delta {\cal F}$ and $\Delta u$.

\subsubsection{Strongly-disordered Medium}

If the tethered system contains a finite concentration of quenched-in
dislocations and/or disclinations, then the density variation $\delta \rho
_0 $ is no longer described by (\ref{drho.cdw}), and one must characterize $%
\delta \rho _0$ statistically through the correlation function $C_0(\br)$ 
or the structure factor $S_0({\bf k})$. Consider the simpler case where 
$S_0({\bf k})$ is liquid-like,  containing no Bragg peaks at finite ${\bf k}$.
This may describe, for example, a completely disordered
film such as a rubber sheet, or a randomly-polymerized liquid membrane.
The model defined by Eqs.~(\ref{W}) and (\ref{H.eff}) 
now contains {\em two} kinds of (mutually uncorrelated) disorders
in the effective random potential $W({\bf r},{\bf u})$, making it difficult
to solve systematically. In fact, straightforward application of the replica
trick to the model (\ref{H.eff}) immediately leads to difficulties as one
has to integrate out the disorders twice.

However, if we simply treat $W({\bf r},{\bf u})$ as an {\em effective} disorder
and examine its moments using (\ref{rho-rho}) and (\ref{V-V}), we find it
has zero mean and a variance given by 
\begin{eqnarray}
\overline{[W(\br,\bu)W(\br',\bu')]}
&=&\Delta _V\delta ^D(\br-\br')\,C_0(\bu-\bu')  \nonumber \\
&\approx&\Delta \,\delta ^D(\br-\br')\,\delta ^D(\bu-\bu'),  \label{W-W}
\end{eqnarray}
with $\Delta =\Delta _VS_0({\bf k}\to 0)$
for short-range correlated function $C_0({\bf r})$.

Note that the form of the correlator~(\ref{W-W}) is the same as that of an
uncorrelated random potential in the space $({\bf r},{\bf u})\in \Re ^{D+D}$. 
It is then tempting to interpret the system (\ref{H.eff}) as a $D$-component, 
$D$-dimensional ``directed manifold'' ${\bf u}({\bf r})$
embedded in an effective $(D+D)$-dimensional random potential 
$W({\bf r},{\bf u})$. The latter is an example of the so-called 
``random manifold'' problem which is encountered in a wide variety 
of context involving quenched randomness~\cite{blatter,frg,thh,lb}. 
Properties of the random manifold (RM) have been characterized
in detail and are briefly summarized here: 
It is known that an $d$-component RM in $D$-dimension is asymptotically
described by a glass phase at any finite temperatures if 
\begin{equation}
2D>d\,(2-D).  \label{D_c}
\end{equation}
Outside this regime, the glass phase is still obtained if the temperature is
below a certain critical temperature~\cite{N4} $T>0$. 
Like the randomly pinned CDW, the behavior of the RM in the glass phase can
be described by the thermal exponent $\theta $ and wandering exponent $\zeta$.
 There is an exact exponent identity $\theta =D-2+2\zeta $ linking the two
exponents for all $d$ and $D$. The exponents are known exactly for the
special case~\cite{dp} $D=1$ and $d=1$, 
with $\zeta =2/3$ and $\theta =1/3$. There are
also strong bounds on the exponents, with $(4-D)/4\leq \zeta \leq \max
\left\{ (4-D)/(4+d),(2-D)/2\right\} $. [Note that $\zeta \rightarrow 0$ as 
$D\rightarrow 4^-$, reflecting the fact that $D=4$ is the upper critical
dimension of the problem.] Numerically, the exponents for various $d$ and $D$
have been determined to good accuracy~\cite{dp,aam}. 
The results are approximately summarized by the expression 
\begin{equation}
\zeta \approx \frac{2(4-D)}{8+d}  \label{zeta.dD}
\end{equation}
which was motivated by a functional renormalization-group 
consideration~\cite{thh}.%

What does the random manifold problem have to do with the problem at hand ?
Even though the second moment of $W({\bf r},{\bf u})$ is of the same form as
that of a $(D+D)$-dimensional random potential, $W$ itself, being the 
{\em product} of two $D$-dimensional random functions, certainly cannot be
truly be a short-range correlated $(D+D)$-dimensional random potential.
There must be long-range correlations, the forms of which are revealed
by considering higher moments of $W$, for example, 
$W_2(\br,\bu) \equiv W^2(\br,\bu) - \overline{\left[W^2\right]}$.
We find 
\begin{equation}
\overline{\left[ W_2(\br,\bu)W_2(0,0)\right]}
= 4 \delta^2(\br) \delta^2(\bu) + 2 \delta^2(\br) 
+2 \delta^2(\br-\bu), \label{W2}
\end{equation}
indicating correlations along the ``directions" of constant $\br$
and constant $\br-\bu$. Such correlations are of course not surprising
given the form of $W(\br,\bu)$ in Eq.~(\ref{W}). 
%By computing higher moments such as 
%$\overline{[W^2({\bf r},{\bf u})\,W^2(0,0)]}$, one immediately
%uncovers the correlated components of the random potential, along the
%directions ${\bf r}={\rm const}$ and ${\bf r}-{\bf u} = {\rm const}$. 
Thus, a more accurate model of the effective random potential should
include a superposition of short-range correlated and long-range 
correlated random potentials, e.g., 
\begin{equation}
W(\br,\bu) = W_0(\br,\bu) + W_a(\br) + W_b(\br-\bu) \label{W0}
\end{equation}
with $W_0(\br,\bu)$ being truly a short-ranged $(D+D)$-dimensional
potential described by the correlator (\ref{W-W}).
The random potentials $W_a$ and $W_b$ {\em generated} from the higher
moments of $W$ have the correlator
$\overline{\left[W_i(\bx)W_j(\bx')\right]} \sim \delta_{ij}
\delta(\bx-\bx')$.

\begin{figure}
\narrowtext
\epsfxsize=2.6truein
\centerline{\epsffile{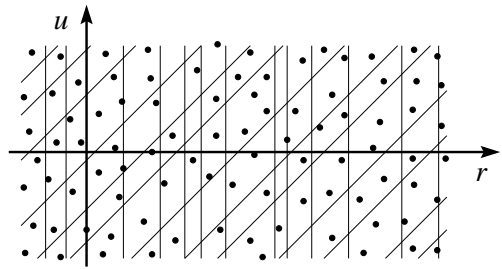}}
\vspace{\baselineskip}
\caption{
Schematic illustration of the effective random potential (\ref{W0}) in $D=1$, 
for a path directed in the horizontal $r$-direction:
The correlated component is shown as two sets of ``columnar defects",
pointing along the directions $r = {\rm const}$ and $r-u = {\rm const}$;
the uncorrelated component is shown as ``point defects". 
}
\label{F02}
\end{figure}

Clearly, $W_a(\br)$ has no effect on the behavior of the system as it
does not involve $\bu$. Since  $|\bu| \ll |\br|$ ($\zeta < 1$),
the $\bu$-dependence in $W_b(\br-\bu)$ is {\em perturbatively irrelevant}, 
and we conclude that $W_b$ does not play a significant role either
in the limit of weak disorders. Another way to gain some intuition of
the correlated potential is through an example in $D=1$. In
Fig.~\ref{F02}, we illustrate the form of the effective random potential
(\ref{W0}) in $D=1$,
where the ``random manifold'' is usually referred to as a ``directed path"
(directed {\em horizontally} in the direction $r$ of Fig.~\ref{F02}), and
the correlated potentials are like the ``columnar defects'' encountered
in the  pinning of flux lines in high-$T_c$ 
superconductors~\cite{nv,hnv,khh,hn}. These columnar defects are
known to be strongly relevant
if oriented in the direction of the directed
path (the $r$-direction in Fig.~\ref{F02}). However, they are irrelevant
if tilted away by a slope exceeding a finite threshold given by the
strength of the disorder.
The correlated potentials $W_a$ and $W_b$  are just
the $D$-dimensional generalization of the ``tilted" columnar defects. 
Their irrelevance can be established more rigorously by extending, for
example, the analysis of Ref.~\cite{hnv} to $D$-dimensions and will
not be pursued here.

%For the problem at hand, the correlated disorders are tilted away from
%the $r$-direction  by slopes of $\sim O(1)$ and $\infty$. 
%They should therefore be irrelevant since $\zeta <1$ implying
%that $|du/dr|\ll 1$ at large scales. The above argument for $D=1$ is easily
%eneralized to $D>1$ as long as $\grad\cdot\bu \ll 1$.

For strong disorders,  a large distortion with $\bu \sim \br$ is admissable
by our model defined by Eqs.~(\ref{H.eff}) and (\ref{W}): 
Such a distortion would be favorable if the elastic 
energy cost (per volume) is more than 
compensated by the disorder energy gained. The latter is given by the
order of variations in $W_b$ and is finite.
Thus, our model can in principle display a phase transition to a ``localized
phase" where $\bu \sim \br$, similar to what was found by Hatano and 
Nelson in the context of non-Hermitian quantum mechanics~\cite{hn}.
However, the formation of this localized phase would require different 
``beads" of the manifold all to lie with a finite volume of the substrate.
This is clearly unphysical and is prevented in practice by any
excluded-volume interaction between the beads.

Based on the above analysis, 
we conjecture that the long-ranged correlations in $W(\br,\bu)$ 
are irrelevant for arbitrary disorder strengths, 
and the problem of a random elastic medium on a
disordered substrate belongs to the same universality class as that of the
random manifold with $d=D$~\cite{ml,N1}. Since the condition (\ref{D_c}) is
always satisfied for $d=D$, we expect our system to be glassy at all finite
temperatures, with the exponents 
\begin{equation}
\zeta \approx \frac{2(4-D)}{8+D}\qquad {\rm and}\qquad \theta \approx \frac{%
D(2+D)}{8+D}  \label{exponents}
\end{equation}
upon adopting the approximate formula (\ref{zeta.dD}).

Our conjecture was presented for the case $D=1$ previously in a short
communication~\cite{prl}. In Sec.~III.B below, we present results of an
extensive numerical study. We find that all universal aspects
of the $D=1$ problem measured, including amplitude ratios, agree
quantitatively with those of the 1+1 dimensional directed path in random
media, thus verifying our conjecture for $D=1$. Very recently, Zeng {\it et
al}~\cite{zeng} applied the min-cut-max-flow algorithm~\cite{aam,mcmf} 
to investigate the $D=2$ version
of the model defined by (\ref{W}) and (\ref{H.eff}), but with only one
component of the displacement field ${\bf u}$. They found an exponent value $%
\zeta \approx 0.42$, which is consistent with that of the random manifold
with $d=1$ and $D=2$ ($\zeta =0.41\pm 0.01$), as expected according to our
conjecture. For a 2-dimensional random elastic medium on a 2-dimensional
substrate, we predict that $\zeta \approx 0.4$ and $\theta \approx 0.8$ as
for a $2$-component, $2$-dimensional directed manifold in $4$-dimensional
random medium. An entangled flux line array, e.g., a ``polymer glass'', 
may be modeled~\cite{drn}  by a 2-component,
3-dimensional random elastic system for time scales up to the distanglement
time $\tau \sim e^{U_{\times }/T}$, where $U_{\times }$ is the flux cutting
energy estimated to be of the order of $10$ times the melting 
temperature~\cite{cut}.
The corresponding exponents are thus expected to be $\zeta \approx 0.2$ and 
$\theta \approx 1.4$ in the elastic regime.

\subsubsection{Nearly-periodic Medium}

We next turn to the case of a nearly-periodic, tethered system with a low
concentration of quenched-in defects. This could be the case of a rigid
vortex array on a thin film superconductor, the defects being frozen-in
vacancies and interstitials~\cite{hmv}. 
Since a low density of vacancies/interstitials
does not destroy the crystallinity of the elastic system, Bragg peaks in 
$S_0({\bf k})$ still leads to a CDW-type interaction (\ref{cdw}) when the
elastic medium is placed in contact with the disordered substrate. However,
the presence of quenched-in vacancies and interstitials also gives rise to
large-scale density variations, manifested by a nonzero component of 
$S_0({\bf k}\rightarrow 0)$, which leads to the effective $(D+D)$-dimensional
random potential as described by the correlator (\ref{W-W}). Thus, the
nearly-periodic elastic system is subject to both the CDW-type and the
RM-type disorders. What is the outcome of competition between these two
kinds of interactions ? A simple power counting reinforces the intuitively
obvious result that the RM
interaction is relevant in the CDW phase, while the CDW interaction is not
relevant in the RM phase. We thereby conclude that even the nearly-periodic
elastic system belongs to the RM universality class.

Our analysis indicates that any finite concentration of quenched-in defects
is sufficient to change the CDW-type pinning of a perfectly periodic system
to the much stronger pinning of the random-manifold universality class. This
includes a low concentration of quenched-in interstitials and vacancies,
which by themselves do not destroy the crystallinity of the elastic medium.
Such a result may be rather surprising at a first glance, since the
existence of periodicity of an elastic medium is usually associated with the
CDW universality class. But this notion is incorrect. What is responsible
for the CDW universality class in the pinning of an usual periodic solid is
a relabeling symmetry of the underlying discrete system, e.g., the energy of
a configuration of beads described by $\left\{ {\bf R}_n\right\} $ is the
same as that described by $\left\{ \bR_{n+{\rm const}}\right\} $ 
(up to boundary effects). This relabeling symmetry is broken~\cite{dsf} 
given a finite concentration of
interstitials/vacancies, since different beads are no longer equivalent.
Thus, asymptotically, the system is controlled by the RM fixed point.

Lastly, we mention that even if the medium contains no topological defects
at all (not even interstitials and vacancies), some quenched-in ``phonon
modes" may already be sufficient to induce the RM behavior: Let the 
equilibrium positions of the beads labeled by $(n_1,n_2)$ be 
$\bR_{n_1,n_2} = \cR_{n_1,n_2} + \bw_{n_1,n_2}$, 
where $\cR_{n_1,n_2}$ denotes the points of a periodic lattice, and
$\bw$ is the quenched-in phonon modes characterized by 
$[(\bw(\bx)-\bw(\bx'))^2]= \Delta (\bx-\bx')$. Then the correlator of the
effective pinning energy $W$ is~\cite{gld1}
\[
\overline{[ W(\br,\bu)W(0,0)]} 
= \Delta_V \, \delta^D(\br) \sum_i \cos({\bf K}_i\cdot\bu)
e^{-\frac{\Delta(\bu)}{2{\bf K}_i^2}},
\]
where ${\bf K}_i$'s are again the reciprocal lattice vectors.
Since $\overline{[ W(\br,\bu)W(0,0)]}$ is
short-ranged correlated in $\bu$ as long as $\Delta (\bu)$ diverges for
large $|\bu|$ (including logarithmic divergence), we see that the
RM universality class is recovered even for quenched-in phonon
fluctuations in $D\le 2$.

\subsection{Transfer Matrix Studies}

We test our predictions by performing numerical studies of a one-dimensional
bead-spring system corresponding to the $D=1$ case of the randomly-tethered
elastic medium considered above. The one dimensional system is chosen since
its thermodynamic properties can be obtained in polynomial ($N^2$) time
using the transfer matrix method~\cite{tm}, 
and also the thermodynamics of the
corresponding $(1+1)$-dimensional problem of a directed path in random media
(DPRM) is known exactly~\cite{dp}. Thus a quantitative comparison can be made.

\begin{figure}
\narrowtext
\epsfxsize=3.0truein
\centerline{\epsffile{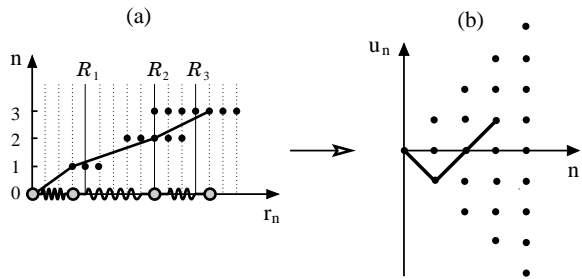}}
\vspace{\baselineskip}
\caption{
Longitudinal displacement variation of the beads of the random chain in (a)
is represented as a directed path on a $(1+1)$-dimensional lattice in (b).
The equilibrium spring lengths are $a_1=4$, $a_2= 5$, and $a_3=3$.
The equilibrium positions $R_n =\sum_{m=1}^n a_m$ are indicated in
(a) as solid lines. The open circles indicate the allowed bead
positions, given the SOS-like restriction on $u_n$.
}
\label{F03}
\end{figure}

We consider the following discretized one-dimensional problem: A chain of 
$N+1$ beads (labeled sequentially by $n\in [0,N]$) is placed on a
one-dimensional lattice of unit lattice spacing. Each bead $n$ (except for $%
n=0$) is connected to its nearest neighbor $n-1$ by a harmonic spring. All
springs have the same spring constant $\gamma $, but the equilibrium length $%
a_n$ is an integer drawn randomly from the interval $[5,15]$ (in units of the lattice
spacing), such that the mean spring length is $a=10$. The equilibrium
positions of the $n^{{\rm th}}$ bead is $R_n=\sum_{m=1}^na_m$ if we fix the $%
n=0$ end of the chain at the origin. 
To speed up the numerics, we apply an SOS-like restriction and allow the
springs to be compressed or stretched by at most one lattice unit. (We have
verified that allowing for large excitations does not affect scaling
properties of a long chain.) Then the chain configuration is given by the
position of each bead $n$ 
\begin{equation}
r_n=R_n+u_n,  \label{E3.1}
\end{equation}
\noindent
where $u_n\in \{0,\pm 1,\pm 2,\ldots \;,\pm n\}$ is the ``displacement
field'', with the restriction $(u_{n+1}-u_n)\in \{0,\pm 1\}$, and the energy
of each spring is $\frac \gamma 2(u_{n+1}-u_n)^2=\left\{ 0,\pm \frac \gamma 2%
\right\} $. Since each chain configuration is uniquely specified by the set
of numbers $\{u_n\}=(u_1,u_2,\ldots ,u_N)$, we may represent the chain as a 
{\em directed path} in $1+1$-dimensions, with ``transverse"
coordinate $\{u_n\}$. 
The mapping is illustrated in Fig.~\ref{F03}. The allowed bead positions
are marked by open circles in Fig.~\ref{F03}(a) and shifted upwards for
clarity. The chain with $N+1$ beads has a total of $3^N$ different
configurations. One of these configuration with $\{u_n\}=(-1,0,1)$ (as shown
in Fig.~\ref{F03}(a)) is represented by the full line in the directed path
representation of Fig.~\ref{F03}(b).

\begin{figure}
\narrowtext
\epsfxsize=2.6truein
\centerline{\epsffile{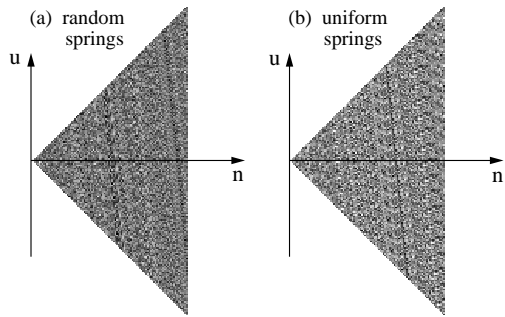}}
\vspace{\baselineskip}
\caption{
Gray-scale plots of potential $W_n(u) = V(R_n + u)$, $R_n = \sum_{m=1}^n
a_m$, for a chain with $N=100$ elements.
(a) Random springs with $a_n$ uniformly distributed in
interval $[5,15]$, and
(b) uniform springs with $a_n=10$.
Lighter shades correspond to larger values of $W$.
}
\label{F04}
\end{figure}

\begin{figure}
\narrowtext
\epsfxsize=2.6truein
\centerline{\epsffile{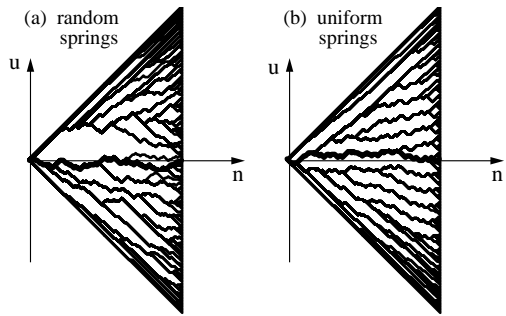}}
\vspace{\baselineskip}
\caption{
Collection of the locally optimal paths for
disorder potential $W$ shown in Figs.~\ref{F04}(a)
and (b). The bold lines represent the global optimal paths.
}
\label{F05}
\end{figure}

Next, we specify the energetics of the system. Disorder of the substrate
with which the chain is in contact is given by the function $V(r)$, drawn
independently from a Gaussian distribution of zero mean and unit variance
for each lattice point $r$. The spring constant used is $\gamma =0.2$ such
that the typical ``spring energy'' is $0.1$, much smaller than the
disordered ``potential energy'' $\sim O(1)$. 
This choice of parameters facilitates a
quick approach to the asymptotic (glassy) regime. In term of the
displacement variable $u_n$, the random potential can be written as $%
W_n(u_n)\equiv V(R_n+u_n)$, which depends explicitly on the two sources of
randomness, $V(r)$ and $R_n$. It is illustrative to plot $W_n(u)$ in the two
dimensional space $(n,u)$ (Fig.~\ref{F04}(a)). For comparison, the case of
uniform springs (with $a_n=10$ for all $n$'s) is shown in Fig.~\ref{F04}(b).
We see that the periodic feature of $W$ shown in Fig.~\ref{F04}(b) is
randomized by the random springs. However, correlations in a slanted
direction is still clearly visible in Fig.~\ref{F04}(a) (cf. Fig.~\ref{F02}).

Effect of the different potential $W$ on the directed path can be
illustrated by differences in the morphology of the ``local optimal paths'',
i.e., the collection of the lowest energy paths~\cite{tm,dp} of length $N$,
connecting the starting point at the origin and all possible ending points
at $(N,u)$. In Figs.~\ref{F05}(a) and (b), we plot the optimal paths for the
realizations of $W$ shown in Figs.~\ref{F04}(a) and (b) respectively. The
bold line corresponds to the global optimal path, the lowest energy path
among all the local optimal paths. Note that the local optimal paths of the
system with uniform springs (Fig.~\ref{F05}(b)) are very regularly arranged,
with the distance between neighboring branches almost constant and
approximately equal to the equilibrium spring length $a$. On the other hand,
the local optimal paths for the random chain are arranged much more
irregularly (Fig.~\ref{F05}(a)). There are for example large islands
of high energy regions which the paths avoid, similar to what was found
previously for the directed path in 2-dimensional random 
potential~\cite{tm,dp}. 
Thus, we can view these paths configurations as an indication of
possible equivalence between the statistical properties of a random chain on
one-dimensional random substrate and a directed path in 2-dimensional
 random medium. 

The statistics of the optimal paths have been investigated in 
Ref.~\onlinecite{prl}, where we presented numerical results obtained from the 
{\em zero-temperature} transfer matrix solution of systems with $N=4096$,
averaged over $2000$ independent realizations of $V(r)$ and $\{a_n\}$ pairs.
Fluctuations in the end-to-end displacement $u_N^{*}$ and the total energy 
$E^{*}(N)$ of the global optimal path $\{u_n^{*}\}$ indicate scaling behavior
with $\overline{[(u_N^{*})^2]} \sim \overline{[u_N^{*}]}
^2\sim N^{4/3}$ and $\overline{[(E_N^{*})^2]} - 
\overline{[E_N^{*}]} ^2\sim N^{2/3}$, both of which are characteristic
of the 1+1 dimensional DPRM universality class~\cite{dp}. 
Here, we describe extension of the
zero-temperature calculations to finite temperatures.

\begin{figure}[tbp]
\narrowtext
\epsfxsize=3.0truein
\centerline{\epsffile{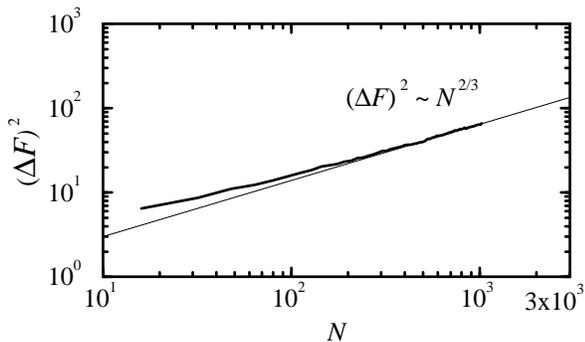}}
\vspace{\baselineskip}
\caption{
Log-log plot of the sample-to-sample free energy fluctuation
$(\Delta {\cal F})^2(N)$ for various chain lengths $N$.
Thin straight line shows the predicted asymptotic scaling.
}
\label{F06}
\end{figure}

\begin{figure}[tbp]
\narrowtext
\epsfxsize=3.0truein
\centerline{\epsffile{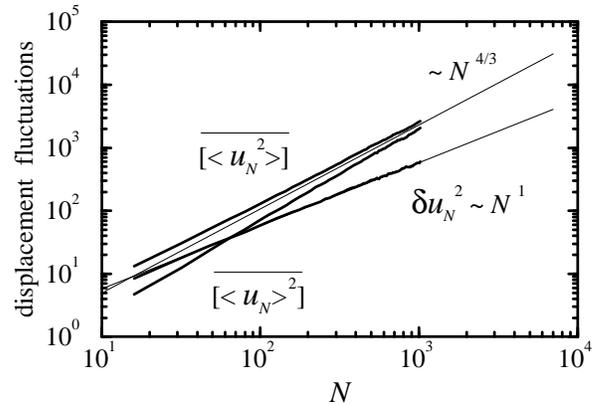}}
\vspace{\baselineskip}
\caption{
Log-log plot of the end-to-end displacement fluctuations for
different chain length $N$, at the temperature $T=1$.
Also shown is the connected thermal fluctuation $\delta u_{N}^{2}$.
}
\label{F07}
\end{figure}

The Boltzmann weight ${\cal W}_N(u)$ for paths connecting their origin
at $(0,0)$ to their end points at $(N,u)$ can be obtained recursively
according to~\cite{tm,tm1}: 
\begin{eqnarray}
{\cal W}_{n+1}(u) &=&c_0(n,u){\cal W}_n(u)  \nonumber \\
&&+c_1(n,u)\Big\{ {\cal W}_n(u-1)+{\cal W}_n(u+1)\Big\},  \label{E3.2}
\end{eqnarray}
\noindent
where $c_0(n,u)=e^{W_n(u)/T}$, $c_1(n,u)=c_0(n,u)\,e^{-\gamma /(2T)}$, with
the ``initial condition'' ${\cal W}_0(u)=\delta _{u,0}$. The  partition
function ${\cal Z}_n$ is obtained as ${\cal Z}_n=\sum_u{\cal W}_n(u)$, 
and the free energy is ${\cal F}_n=-T\ln {\cal Z}_n$. The thermal
averaging of an observable ${\cal O}_n(u)$ is given by 
\begin{equation}
\langle {\cal O}_n\rangle \equiv \frac{\sum_u{\cal O}_n(u){\cal W}_n(u)}{%
\sum_u{\cal W}_n(u)}.  \label{E3.3}
\end{equation}

The sample-to-sample free energy fluctuation 
$(\Delta {\cal F})^2= 
\overline{[{\cal F}_N^2]} -\overline{[{\cal F}_N]}^2$ is
shown in Fig.~\ref{F06}. The results are obtained at temperature $T=1.0$,
from systems with $N=1024$ averaged over 4000 different realizations of
random $V(r)$ and $\{a_n\}$. It is seen that $(\Delta {\cal F})^2(N)$
approaches the asymptotic scaling form $N^{2/3}$. We also computed the
finite-temperature end-to-end displacement fluctuations $\overline{
[\langle u_N^2\rangle ]}$ and $\overline{[\langle u_N\rangle ^2]}$.
 As shown in Fig.~\ref{F07}, the approach to the expected $N^{4/3}$
behavior is clear. More interestingly, note that the disorder average of the
thermal fluctuation itself, i.e., the connected second moment $\delta
u_N^2\equiv \overline{[\langle u_N^2\rangle -\langle u_N\rangle ^2]}$
in fact scales as $N^1$ as if randomness is not present. 
This behavior is
expected of the DPRM due to a statistical tilt symmetry~\cite{dp,gi,fh}. 
Such a symmetry is
obviously not present in the bare potential $W_n(u)$ shown in 
Fig.~\ref{F04}(a). The scaling law on $\delta u_N^2$ found is 
therefore another evidence
indicating the irrelevancy of the slanted correlation in $W_n(u)$ and the
dominance of the 1+1 dimensional DPRM behavior. 

\begin{figure}
\narrowtext
\epsfxsize=3.0truein
\centerline{\epsffile{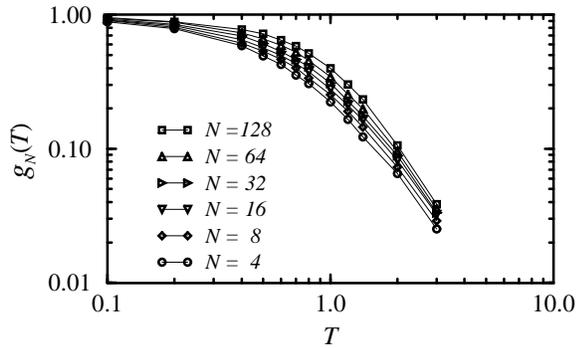}}
\vspace{\baselineskip}
\caption{
Log-log plot of $g_N(T)$ as a function of temperature,
averaged  over $10^5$ different realizations of disorder.
The high temperature behavior approaches $1/T^2$, which is
expected of the $1+1$ dimensional DPRM.
}
\label{F08}
\end{figure}

\begin{figure}
\narrowtext
\epsfxsize=3.0truein
\centerline{\epsffile{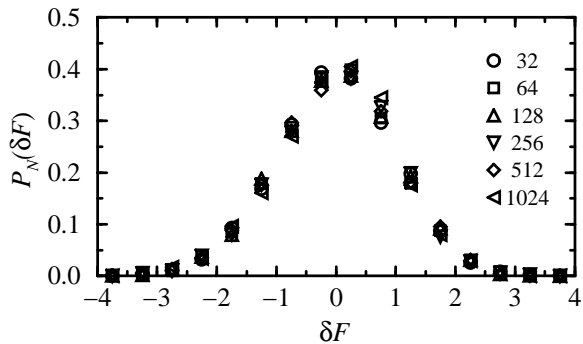}}
\vspace{\baselineskip}
\caption{
Probability distribution for the free energy, plotted for the
rescaled variable,
$\delta {\cal F} = \left({\cal F} - \overline{[{\cal F}]}\right)
/\Delta {\cal F}(N)$, at $T=1.0$
for system sizes ranging from $N=32$ to $N=1024$. Data were
collected from $4096$ samples.
}
\label{F09}
\end{figure}

Our results suggest that the finite temperature behavior of the system
probed is dominated by the $T=0$ fixed point. This result is not immediately
generalizable to all $T$. Naively, one might think that at sufficiently high 
$T$ such that the thermal fluctuation $\delta u_N$ exceeds the quenched
variation in $R_N$, then the effect of random spring length may be washed
out. To investigate the possible existence of a finite temperature phase
transition, we compute the dimensionless quantity~\cite{kmb} 
\begin{equation}
g_N(T)=\overline{[\langle u_N\rangle ^2]} \left/ 
\overline{[\langle u_N^2\rangle ]}\right. ,  \label{E3.3a}
\end{equation}
which vanishes for $T\rightarrow \infty $ while $g\rightarrow O(1)$ for 
$T\rightarrow 0$. If there is a phase transition at some finite temperature 
$T_c$, the expected finite-size scaling in the critical region would be 
$g_N(T)=\widetilde{g}\big( N^{1/\nu }(T-T_c)\big) $, where $\nu $ is the
correlation length exponent. Therefore, the curves of $g(T)$ for different 
$N $'s should all intersect each other at $T_c$ if a finite temperature
transition exists. 
The numerical data for $g_N(T)$ in the temperature range $T=0.1$ to $3.0$, 
for system sizes from $N=4$ to $N=128$ are shown in Fig.~\ref{F08}. No 
indication of curve crossing at $T>0$ is found. 
In fact, the high temperature behavior of $g_N(T)$ approaches $1/T^2$
similar to that of the $(1+1)$-dimensional DPRM which has no finite
temperature phase transition~\cite{kmb}. We therefore
conclude that in the thermodynamic limit $N\rightarrow \infty $, there is no
finite temperature phase transition, and the large scale behavior of
the random chain is always described by the zero-temperature (glassy) system.

\begin{figure}
\narrowtext
\epsfxsize=3.0truein
\centerline{\epsffile{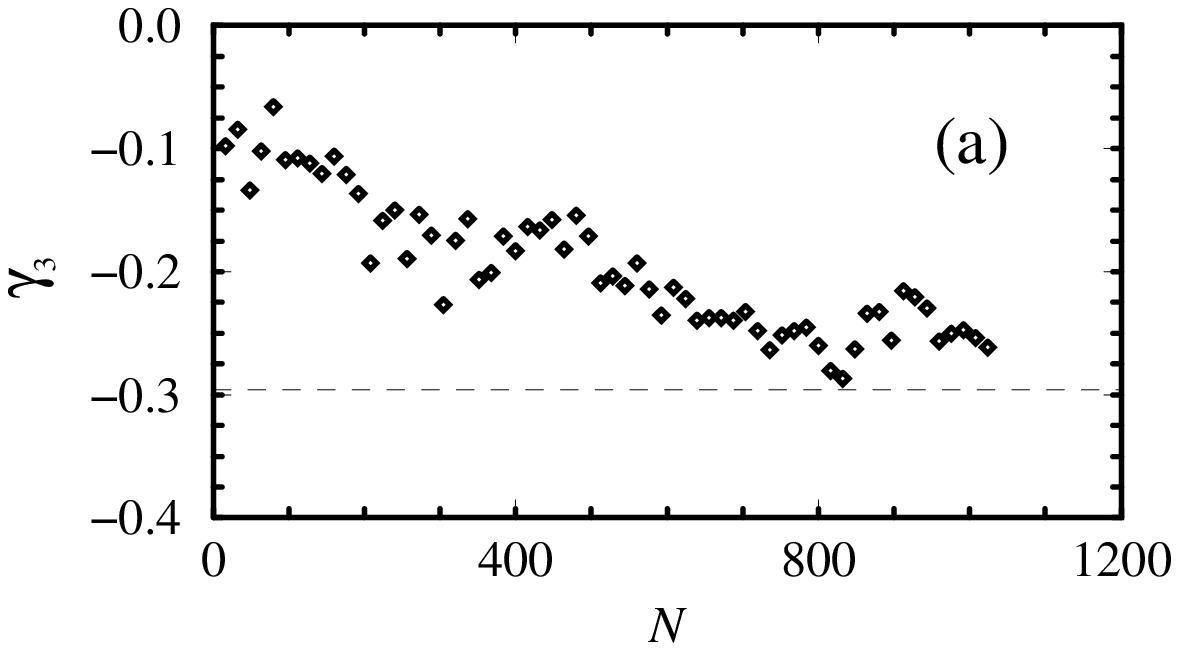}}
\vspace{\baselineskip}
\epsfxsize=3.0truein
\centerline{\epsffile{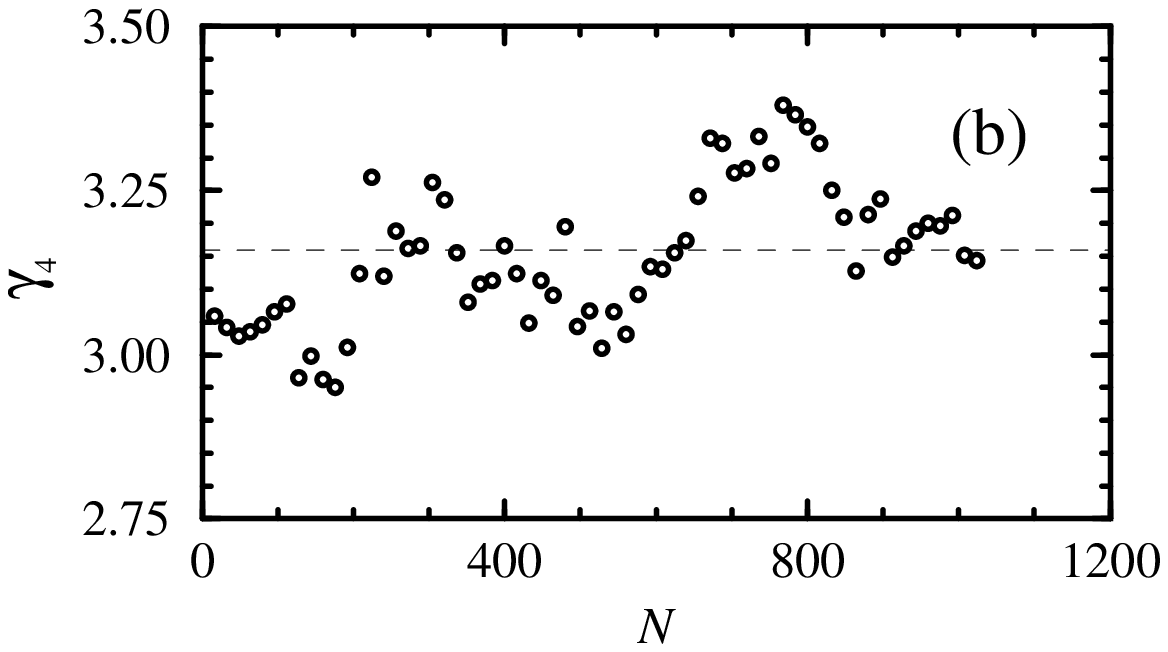}}
\vspace{\baselineskip}
\caption{
Skewness $\gamma_3$ and kurtosis $\gamma_4$ of the free energy
 distribution for different chain lengths $N$.
Dashed lines are the corresponding values for the $(1+1)$-dimensional
DPRM: $\gamma_3 \approx -0.296$ and $\gamma_4 \approx 3.16$.
}
\label{F10}
\end{figure}

We have also calculated the full free energy probability distribution $P$
for different lengths $N$. Figure~\ref{F09} shows data collapse of 
$P_N(\delta {\cal F})$ for $N=32$ to $N=1024$, collected from 4096 samples,
with $\delta {\cal F}\equiv ({\cal F}-\overline{[{\cal F}]})
/\Delta {\cal F}(N)$ being the dimensionless measure of free energy
variation. 
The distribution is asymmetric. The skewness 
$\gamma _3$ and kurtosis $\gamma _4$ of this distribution are plotted in 
Fig.~\ref{F10}(a) and Fig.~\ref{F10}(b) respectively for different $N$'s.
Both $\gamma_3$ and $\gamma_4$ are found to approach 
the corresponding  universal numbers (dashed lines) known for 
the (1+1)-dimensional DPRM~\cite{kmb}: 
$\gamma_3=-0.296\pm 0.028$, $\gamma _4=3.16$. 

Putting together all the numerical results presented in this section, we see
strong evidence supporting our expectation that the thermodynamics of a
random chain on the disordered substrate and the (1+1)-dimensional DPRM
indeed belong to the same universality class.

\section{Driven Dynamics}

\subsection{Theoretical Considerations}

\subsubsection{Driven CDW and Random Manifolds}

The zero-temperature driven dynamics of the tethered system is of interest
to the study of tribology and to understanding the nonequilibrium dynamics
of  vortices. From the
correspondence between the {\em thermodynamic} properties of the
randomly-tethered elastic system and the CDW/RM systems, it is tempting to
conjecture that the correspondence persists also for the dynamical
properties. Before we provide evidences in support of this generalization,
let us first review the known dynamics of the driven CDW/RM systems.

Consider first the simplest zero-temperature driven dynamics of the
randomly-pinned (one-component) CDW in $D$-dimension~\cite{cdw1,nf}, 
given by the equation of motion 
\begin{equation}
\mu _0^{-1}\partial _t u(\br,t)=-\frac \delta {\delta u}
{\cal H}_{{\rm CDW}}\{ u\} +F  \label{e.cdw}
\end{equation}
where $\mu _0$ is a bare frictional coefficient, and $F$ is the driving
force. For $F$ below some threshold force $F_c$, the average velocity 
$v\equiv \left\langle \partial _tu\right\rangle $ is zero. 
[Here, $\left\langle ...\right\rangle $ denotes temporal and spatial average.]
Upon approaching the threshold from below, the dynamics (e.g., response to
perturbation) becomes very ``jerky''. It is consisted of a series of
``avalanches'', whose (linear) size $\ell $ obeys a power-law 
distribution~\cite{nm}, 
\begin{equation}
\Pr (\ell >s)=s^{-\kappa }\,\widehat{\rho }(s/\xi ).  \label{avalanche}
\end{equation}
In Eq.~(\ref{avalanche}), $\xi $ is the correlation length of the system, 
$\widehat{\rho }(x)$ is a scaling function which is constant for $x\ll 1$ and
drops off sharply for $x\gg 1$. The correlation length diverges as $\xi \sim
(F_c-F)^{-\nu }$ as $F\rightarrow F_c^{-}$. The motion becomes continuous
for $F>F_c$ due to overlapping avalanches. There the interface acquires a
finite velocity with $v\sim (F-F_c)^\beta $, similar to the emergence of the
order parameter in a critical phenomenon. These exponents have been computed
by a functional renormalization group (FRG) analysis~\cite{nf} 
to first order in 
$\epsilon =4-D$, with $\nu =1/2$ and $\beta =1-\epsilon /6$. The one-loop FRG
results are found to be consistent with extensive numerical simulations of
the driven CDW in various spatial dimensions~\cite{pbl,mf,myers}. 
For example, Myers and Sethna~\cite{myers} find the
one-dimensional driven CDW ($\epsilon =3$)  to have
 $\nu \approx 0.4\pm 0.1$ and $\beta \approx 0.45\pm 0.05$.

Similar depinning phenomenon~\cite{ke1} 
is obtained for the zero-temperature driven
dynamics of the $d$-component random manifold ${\bf u}({\bf r)}$, whose
equation of motion is 
\begin{equation}
\mu _0^{-1}\partial _t{\bf u}({\bf r},t)=-\frac \delta {\delta {\bf u}}{\cal %
H}_{{\rm RM}}\left\{ {\bf u}\right\} +{\bf F},  \label{e.rm}
\end{equation}
where ${\cal H}_{{\rm RM}}$ refers the random manifold 
Hamiltonian~(\ref{H.eff}) with a truly random $(d+D)$-dimensional potential 
$W$, and 
${\bf F}$ is again the driving force. A continuous depinning transition
occurs at a critical force $\left| {\bf F}\right| =F_c$, where the system
exhibits avalanches with power-law distribution as in (\ref{avalanche}). In
the vicinity of the critical point, the exponents $\nu $ and $\beta $ can be
defined as before. [Here, $\beta $ is the exponent describing the onset of
the parallel component of the velocity, $v\equiv \left\langle \partial _t%
{\bf u}\right\rangle \cdot {\bf \hat{F}}$.] As the driving force ${\bf F}$
breaks the isotropy of the system, it is convenient~\cite{ke2} to divide the
displacement ${\bf u}$ into components parallel and perpendicular to $\bF$,
 with $u_{\parallel }\equiv {\bf u}\cdot {\bf \hat{F}}$ and 
${\bf u}_{\perp }\equiv {\bf u}-u_{\parallel }\,{\bf \hat{F}}$. The depinning
transition can then be understood in terms of the critical fluctuations in 
$u_{\parallel }$ and ${\bf u}_{\perp }$, given by the correlation functions 
\begin{eqnarray}
\left\langle \big( u_{\parallel }({\bf r},t)
-u_{\parallel }({\bf 0},0)\big)^2\right\rangle  
&=&|{\bf r}|^{2\chi _{\parallel }}\,\hat{g}_{\parallel}
(|t|/|{\bf r}|^{z_{\parallel }})  \label{corr.1} \\
\left\langle \big({\bf u}_{\perp }({\bf r},t)
-{\bf u}_{\perp }({\bf 0},0)\big)^2\right\rangle  
&=&|{\bf r}|^{2\chi _{\perp }}\,
\widehat{g}_{\perp}(|t|/|{\bf r}|^{z_{\perp }})  \label{corr.2}
\end{eqnarray}
at $\left| {\bf F}\right| =F_c$. They are characterized by their respective
roughness exponent $\chi _{\parallel ,\perp }$ and dynamic exponent 
$z_{\parallel ,\perp }$, from which all other exponents can be 
obtained~\cite{ke1}. For example, $\nu =1/(2-\chi _{\parallel })$, 
$\beta =\nu \,(z_{\parallel }-\chi_{\parallel })$ 
and $\kappa =D-2+\chi _{\parallel }$~\cite{exp}. It was shown by Ertas and 
Kardar~\cite{ke2} that the scaling properties of $u_{\parallel }$ are 
the same as
those of the one-component system (RM with $d=1$) which has been solved by
the FRG method~\cite{natt,nf2}, with $\chi _{\parallel }=\epsilon /3$ and 
$z_{\parallel }=2-2\epsilon /9$ to first order in $\epsilon =4-D$. Ertas and
Kardar also found~\cite{ke2} $\chi _{\perp }=\chi _{\parallel }-\frac D2$ 
and $z_{\perp}=z_{\parallel }+\nu ^{-1}$.

Numerically, one finds for the one-dimensional interface in 
(1+1)-dimensions~\cite{string}
(i.e, $d=D=1$ or $\epsilon =3$) that $\chi \approx 0.97\pm 0.05$, $\nu
\approx 1.05\pm 0.1$, and $\beta \approx 0.24\pm 0.1$. For the
two-dimensional interface in 3-dimensions~\cite{2drm} 
($d=1$ and $\epsilon =2$), the
exponents are $\chi \approx 0.67\pm 0.03$ and $\nu \approx 0.75\pm 0.05$.
These results are all consistent with the one-loop 
FRG predictions~\cite{natt,nf2}. 
In addition, a different exponent is found for the finite size scaling
of the roughness~\cite{lt}, with
$$
\langle \big(u(L,t)-u(0,t)\big)^2\rangle \sim L^{2\chi_{\rm FS}},
$$
with $\chi _{{\rm FS}}\approx 1.25$ in $D=1$.

\subsubsection{Driven Tethered Network}

We now return to the randomly tethered elastic network defined in Sec.~II,
and study its motion in the presence of a uniform driving force ${\bf F}$
parallel to the substrate. We start with the deterministic and purely
dissipative dynamics of the discrete bead-spring system. The equation of
motion, in terms of the displacement vector ${\bf u}({\bf R}_n)$
for the bead $n$, has the form 
\begin{eqnarray}
\mu _0^{-1}\partial _t{\bf u}({\bf R}_n,t) &=&-\sum_{\{n^{\prime
}\}}\mG^{-1}({\bf R}_n-{\bf R}_{n^{\prime }})\cdot {\bf u}(%
{\bf R}_{n^{\prime }},t)  \nonumber \\
&&-\frac \delta {\delta {\bf u}}V[{\bf R}_n+{\bf u}(%
{\bf R}_n,t)]+{\bf F},  \label{E2.7}
\end{eqnarray}
where the kernel $\mG^{-1}$ describes the spring forces exerted by all
the beads $\{n^{\prime }\}$ connected with the bead $n$.

As we will show in Sec.~IV.B, the behavior of this tethered network is very
similar to that of the driven CDW and RM just described: At small driving
forces the system is completely pinned by the random potential. As the force
increases above some threshold value $F_c$, the system starts to move with
a nonzero average velocity, $v\sim (F-F_c)^\beta $.
The behavior near the depinning transition is characterized by
a diverging correlation length $\xi \gg a$ as in usual critical phenomena.

To study  the depinning phenomenon, it is useful to {\em coarse grain }
the equation of motion for the discrete system (\ref{E2.7}).
The procedure is described in  Appendix A 
for a 2D system.
In term of the coarse-grained displacement field ${\bf u}$, the
equation of motion becomes
\begin{equation}
\mu _0^{-1}\partial _t{\bf u}({\bf r},t) = (1-\grad\cdot \bu) 
\tf(\br,\{\bu\}) + \fpin(\br,\bu)
\label{eom.u}
\end{equation}
where
\begin{equation}
\tf(\br,\{\bu\})=-\int_{\br'} 
\mG^{-1}({\bf r}-{\bf r^{\prime }})\cdot {\bf u}({\bf r}^{\prime },t) 
- \grad V_<(\br) + \bF
\label{bbf}
\end{equation}
results from the straightforward coarse graining of the right-hand side
of Eq.~(\ref{E2.7}), with $V_<(\br)$ being the slowly varying part of
the substrate potential $V(\br)$.
Since the beads in the tethered systems are connected only to other beads in
their vicinities, the kernel $\mG^{-1}$ is local. Its Fourier transform
reads in component form 
\begin{equation}
\widehat{G}_{ij}^{-1}\left( {\bf k}\right) =c_{66}\delta
_{ij}k^2+(c_{11}-c_{66})k_ik_j,  \label{G.inv}
\end{equation}
where we have again neglected the spatial variations in the elastic moduli.
Finally, the ${\bf u}$-dependent pinning force in (\ref{eom.u}) is 
\begin{equation}
\fpin({\bf r},{\bf u})=-\grad V_>(\br) \cdot
\delta \rho _0({\bf r}-{\bf u})/\bar{\rho},  \label{E2.9}
\end{equation}
where $V_>(\br)$ are the Fourier modes of $V(\br)$ close to the inverse
bead spacing.

For the system with uniform springs, density variation is given by 
(\ref{drho.cdw}). The equations of motion (\ref{eom.u}) -- (\ref{E2.9}) 
are then
similar to the ones describing the randomly-pinned ($D$-component) CDW. This
is expected given the thermodynamic properties described in Sec.~III. 
Compared to the ``usual'' equation of motion for the driven CDW, 
$\mu_0^{-1} \partial_t \bu = - \delta {\cal H}_{\rm CDW}/\delta\bu + {\bf F}$, 
Eqs.~(\ref{eom.u}) -- (\ref{E2.9}) contain addition
terms such as  $\grad V$ and $ (\grad \cdot \bu) \, \tf$.
These terms  have recently been introduced on phenomenological 
grounds~\cite{R19,R20}. 
Here we find that they can be obtained systematically~\cite{gld2} from
a coarse-graining procedure (Appendix A). What effects do these additional
terms have in the vicinity of the depinning transition ? 
The term $\grad V$ is ${\bf u}$-independent and hence does
not provide pinning~\cite{gradv}. The
term proportional to $(\grad \cdot \bu)$ is dynamic in origin. 
Kinetically-generated
terms of this kind, including other terms 
with higher powers in $(\grad \cdot \bu)$,
can drastically affect the behavior of the system in the
limit of strong drive where $\tf$ is large.
However, they are irrelevant
in the vicinity of the depinning transition 
where $\partial_t \bu,\tf \to 0$~\cite{aniso}.

The introduction of random springs destroys periodicity in $\delta \rho _0$
and one must describe the random pinning force 
$\fpin({\bf r},{\bf u})$
statistically through the correlation function $C_0$ as done in the
equilibrium case. We find 
\begin{eqnarray}
\overline{[f_{{\rm pin},i}({\bf r},{\bf u})
f_{{\rm pin},j}(0,0)]}
&=&\frac{\Delta_V}{ \brho^2} \, \left(\frac{2\pi}{a}\right)^2 \delta_{ij}\, 
\delta (\br) C_0(\bu), \nonumber \\
&\approx& \frac{\Delta_V}{\brho^2} \delta (\br)\partial _i\partial_j C_0(\bu)  
\label{f-f}
\end{eqnarray}
which is short-range correlated in both $\br$ and $\bu$
for short-ranged correlated  $\delta \rho _0$.
As in the static case analyzed in Sec.~III, higher moments of 
$\fpin$ are long-range correlated but  irrelevant.
The critical dynamics with short-ranged pinning forces is then  
in the  universality class of the driven RM~\cite{nf2}.
In fact,  the pinning force appears as if generated directly from the
effective potential $W(\br,\bu)$, i.e., 
$\fpin \approx - \delta W(\br,\bu)/\delta \bu$. We thus
conjecture that the critical depinning dynamics of the $D$-dimensional
randomly-tethered elastic network on disordered substrate is in the same
universality class as the $D$-component, $D$-dimensional directed manifold
in $(D+D)$-dimensional random medium~\cite{N5}. 

\subsection{Numerical Simulations}

In this section, we report detailed numerical simulation of the critical
dynamics of the driven one-dimensional bead-spring system whose equilibrium
properties were described in Sec.~III.B. The zero-temperature response of the
random chain to an external driving force is obtained by a direct numerical
integration of the overdamped equation of motion, 
\begin{eqnarray}
\partial _tr_n(t) &=&\gamma \left[ r_{n+1}(t)-2r_n(t)+r_{n-1}(t)\right. 
\nonumber \\
&&\quad \quad \left. -(a_{n+1}-a_n)\right] - V'(r_n)+F  \label{E3.4}
\end{eqnarray}
where $r_n$ is the position of the $n^{{\rm th}}$ bead, $a_n$ is the
equilibrium length of the $n^{th}$ spring and $V'=-dV/dr$. Expressed in
term of the displacement field $u_n(t)=r_n(t)-R_n$, where 
$R_n=\sum_{m=1}^na_m$, Eq.~(\ref{E3.4}) is just the one-dimensional version
of Eq.~(\ref{E2.7}). (The bare frictional coefficient $\mu _0$ is set to unity
here.) We integrate Eq.~(\ref{E3.4}) in the simple Eulerian manner by
discretizing time. The positions $r_n$'s are kept as continuous variables. 
The random potential $V(r)$ is constructed by a series of
(quadratic) bumps and valleys, centered on a lattice with unit spacing,
i.e., at $j=\{0,\pm 1,\pm 2,\ldots \}$. The range of each bump/valley is
$R_0\leq 1/2$, and the amplitude $V_j$ of the bump/valley centered at site $j$ is drawnly randomly from the interval $[-1,1]$. Thus, 
\begin{equation}
V(r)=\sum_j\frac{V_j}{2}\big[(r-j)^2-R_{0}^{2}\big]
        \Theta\big(R_0-|r-j|\big).
\label{E3.5}
\end{equation}

Since in Eq.~(\ref{E3.4}) there is no term which forbids local
crossings among the
chain elements, in the evolution of $r_n(t)$ we explicitly put additional
restriction eliminating such moves, so that
$r_{n+1}\ge r_n$ always holds. This speeds up the dynamics and does not
affect the asymptotic scaling behavior as we verified. 
The simulations were run on
different system sizes with $\gamma =0.1$, $R_0=1/2$,
and spring lengths $\{a_n\}$ chosen randomly from the interval $[5,15]$. 
Various time step sizes were
used, ranging from $\Delta t=0.5$ to $\Delta t=0.01$. For all the results
reported, we always checked that twice smaller $\Delta t$ does not lead to
significant differences.  
Open chain boundary conditions were imposed by introducing two fictitious
beads, with
 $r_{-1}(t)\equiv r_{0}(t)-a_0$, and $r_{N+1}(t) \equiv r_N(t)+a_{N+1}$
supplementing Eq.~(\ref{E3.4}).
As initial conditions, we take each spring to be
either compressed or stretched, within $50\%$ of its equilibrium length. The
chain is then released in the random environment described by Eqs.~(\ref
{E3.4}) and (\ref{E3.5}). Each bead is pulled by a constant force $F$ which
is the only parameter we vary.

\begin{figure}
\narrowtext
\epsfxsize=3.0truein
\centerline{\epsffile{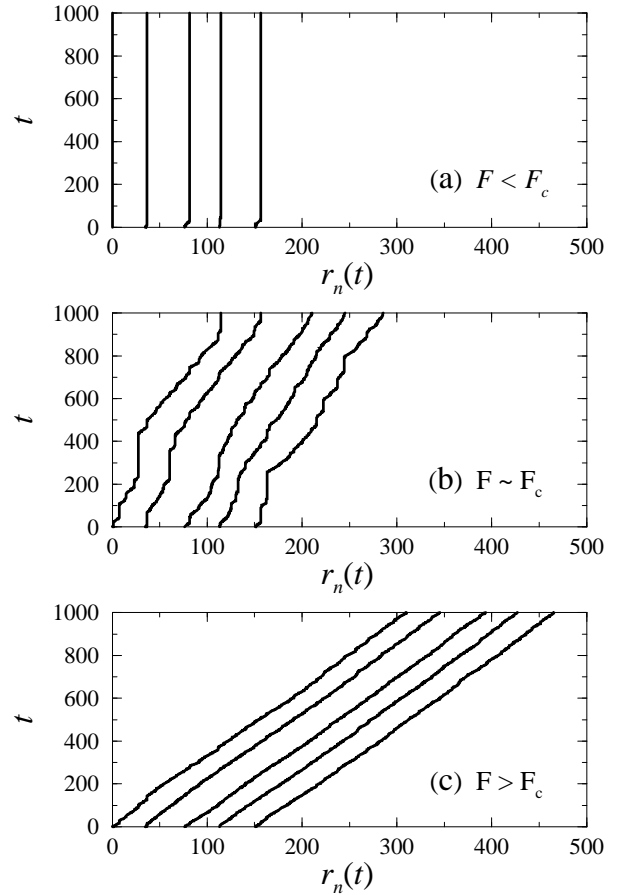}}
\vspace{\baselineskip}
\caption{
Trajectories of a system of $N=16$ beads
for three characteristic driving forces $F$.
Size of time step is $\Delta t = 0.01$.
}
\label{F11}
\end{figure}

By applying a strong enough force, the system starts to move with a velocity
which after some initial fluctuation, reaches its stationary value 
\begin{equation}
v(F)\equiv \frac 1N\sum_{n=1}^N\overline{[\langle \partial
_tr_n\rangle ]} .  \label{E3.6b}
\end{equation}
Here $\langle \cdots \rangle $ represents averaging over long time, which is
very much needed in the vicinity of the depinning transition where the chain
motion becomes very jerky as we illustrate below: In Fig.~\ref{F11}, we show
the bead trajectories $r_n(t)$ for a chain with $N=16$ beads. For clarity,
only the trajectories of beads with indices $n=0,4,8,12,16$ are shown. The
driving force is $F=0.2$ in Fig.~\ref{F11}(a). After some initial movements,
all the trajectories become independent of time, indicating that the chain
is pinned to one of its meta-stable states. In Fig.~\ref{F11}(b) the same
chain is driven by a stronger force, $F=0.29$. The dynamics is characterized
by jerky, non-uniform motion. In the time interval monitored, the chain
moved very slowly over a distance of the order of its length. Increasing the
force further, the trajectories become smooth again and are little affected
by disorder. Figure~\ref{F11}(c) shows an example with $F=0.4$. The beads
march forward with a finite velocity which is given by the slopes of the
trajectories. By measuring the average slopes for different $F$'s, we
determine the velocity-force characteristics $v(F)$. 

\begin{figure}
\narrowtext
\epsfxsize=3.0truein
\centerline{\epsffile{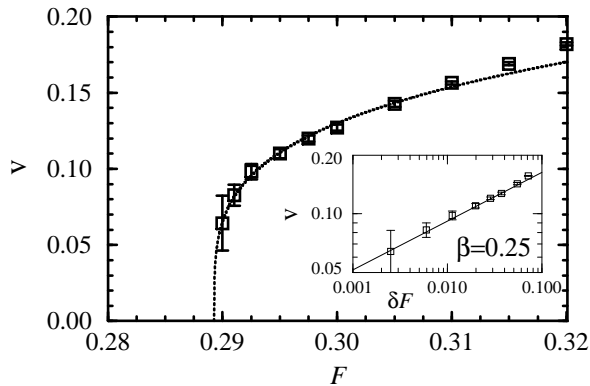}}
\vspace{\baselineskip}
\caption{
Average velocity $v$ as a function of the force $F$ exerted
on the random spring chain. Dotted
line is the best fit to the form $v=v_0(F-F_c)^{\beta}
$ where constants $v_0$, $F_c$, and $\beta$ are fitting parameters.
Inset shows a log-log plot of $v$ vs the reduced force 
$\delta F=(F-F_c)/F_c$;
straight line indicates the suggested scaling behavior.
}
\label{F12}
\end{figure}

\begin{figure}
\narrowtext
\epsfxsize=3.0truein
\centerline{\epsffile{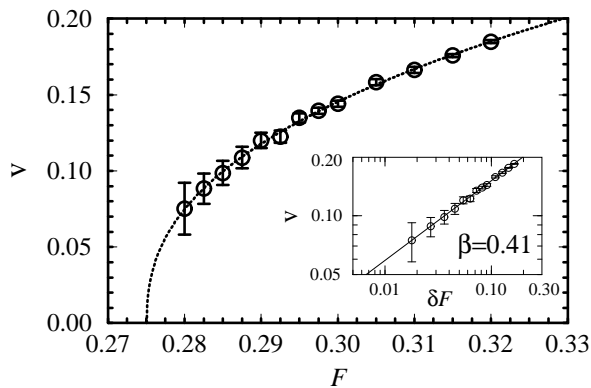}}
\vspace{\baselineskip}
\caption{
The $v-F$ characteristics for a {\em uniform} spring
chain. Dotted and straight lines are the best fit curves
analogous to the ones shown in Fig.~\ref{F12}.
}
\label{F13}
\end{figure}

Numerical results for $v(F)$ are shown in Fig.~\ref{F12} for systems of size 
$N=1024$. Depending on the value of $F$, the time averages were taken over
the intervals of $10^5$ to $10^7$ steps. Further disorder average was
performed over 10 to 50 different realizations of $\{V_j\}$ and $\{a_n\}$.
(For $F$ far exceeding $F_c$, there is no need for large number of samples.)
The data clearly indicated a sharp rise in $v$ at a threshold
force of $F_c\approx 0.289$. In the inset of Fig.~\ref{F12}, 
we plot $v$ against the
reduced driving force $\delta F\equiv (F-F_c)/F_c$ on log-log scale. This
yields the expected scaling form, $v\sim \left( \delta F\right) ^\beta $,
with $\beta \approx 0.25$. For comparison, we also show the corresponding $%
v-F$ characteristics for the case of uniform springs (with $a_n=10$) in
Figs.~\ref{F13}. We again find critical depinning behavior, with $%
F_c\approx 0.275$ and $\beta \simeq 0.41$.

Figs.~\ref{F12} and \ref{F13} clearly show the difference between the
uniform and random spring systems. In the case of uniform springs, our
estimate for exponent $\beta $ is comparable with the result $\beta =0.45\pm
0.05$ obtained by Myers and Sethna~\cite{myers} in their simulations of a
one-dimensional automation model believed to be in the CDW universality
class; it is also consistent with the FRG result~\cite{nf} for the 1D CDW 
($\beta =1/2$) described in Sec.~IV.A. For the random chain, 
our value of $\beta $ is
close to the one obtained in numerical studies of the dynamics of a directed
elastic string in $2-$dimensional random media~\cite{string}, 
which finds $\beta
=0.24\pm 0.1$ for strong pinning and $\beta =0.34\pm 0.1$ for weak pinning.

\begin{figure}
\narrowtext
\epsfxsize=3.0truein
\centerline{\epsffile{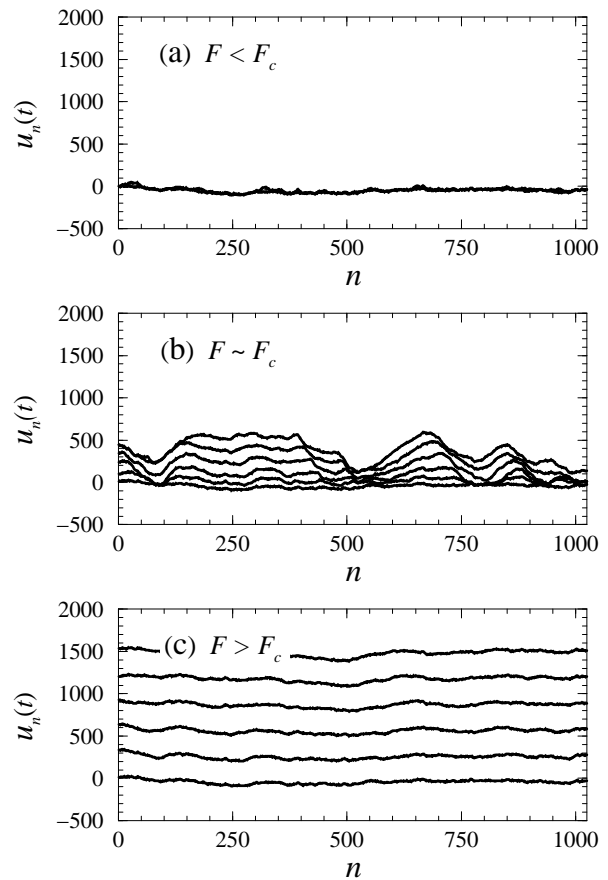}}
\vspace{\baselineskip}
\caption{
Temporal evolution of the displacement profile for the
random chain of $N=1024$ beads at different driving forces $F$.
The lines plotted are displacement profiles at $t=0,1000,
\ldots, 5000$, with time step of $\Delta t = 0.01$.
}
\label{F14}
\end{figure}

Next, we characterize fluctuations in the bead configurations in the
vicinity of the critical point. Positional fluctuations implicit in 
Figs.~\ref{F11} can be visualized more directly by plotting the displacement 
field $u_n(t)=r_n(t)-R_n$. 
In Fig.~\ref{F14}, we show the typical response of the random chain with 
$N=1024$ beads to the applied force in the three regimes below/at/above the
depinning transition, all with the same realization of disorders. For $F<F_c$
(Fig.~\ref{F14}(a)), all beads stop to move after some transient motion
characterized by small, local rearrangements. 
For $F\approx F_c$, the motion of the system is highly nonuniform as shown
in Fig.~\ref{F14}(b). (Here, the lines plotted are the displacement profiles
for $t=0,1000,\ldots ,5000$ with $\Delta t=0.01$.) 
The profile advances in a very jerky fashion
reminiscent of avalanches observed in models of sandpile and 
earthquakes~\cite{cbo,btw}.
Note that these avalanches have sizes comparable to the system size, making
the displacement profiles much ``rougher'' than the ones shown in 
Fig.~\ref {F14}(a). 
For $F>F_c$, the avalanches {\em overlap} each other and the displacement
profile cannot stop moving as they did in the previous cases~\cite{overlap}. 
Fig.~\ref{F14}(c) shows the system evolution when the driving force is much
larger than $F_c$. (The chain positions are plotted at a time interval of 
$t=1000$ as in Fig.~\ref{F14}(b)). Since this time interval much exceeds the
avalanche overlap time, individual avalanches are not discernible at this
scale, and the system advances smoothly with a reduced roughness (governed
by the KPZ equation~\cite{kpz}).

We now characterize the roughness of the displacement profile
quantitatively close to the critical point. 
We monitor the disorder-averaged equal-time correlation
function, 
\begin{equation}
C(n,N)=\frac 1{N}\sum_m \overline{\big[\big(r_{m+n}(t)-r_m(t) 
- n\;a\big)^2\big]}.  
\label{E3.7}
\end{equation}
Systems with $N=8$ to $256$ beads were examined right at the
threshold forces $F_c$. Simulations were run until the systems become
``barely pinned'', defined operationally as the point where  $v<10^{-3}$. 
Disorder averages were performed
over  $5000$ samples for the smaller $N$'s and $100$ samples for
the largest $N$.

\begin{figure}
\narrowtext
\epsfxsize=3.0truein
\centerline{\epsffile{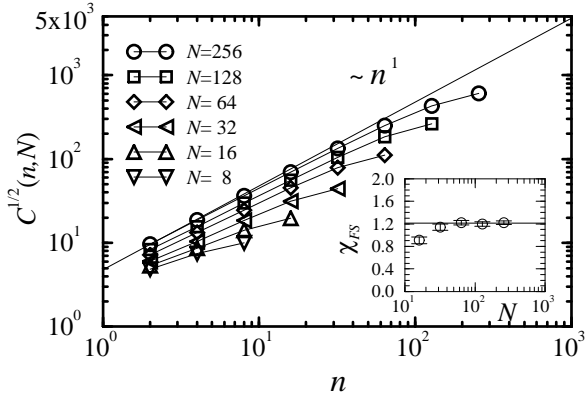}}
\vspace{\baselineskip}
\caption{
Correlation functions for random spring chains of size $N$.
The straight line shows $C(n,N) \sim |n|^1$.
Inset shows the effective finite-size roughness exponent
$\chi_{\rm FS}(N)$, defined by Eq.~(\ref{E3.8}).
Dotted line is the suggested asymptotic value of the exponent.
}
\label{F15}
\end{figure}

Numerical results for $C(n,N)$ are shown in Fig.~\ref{F15} for the random
chain. Because of the SOS-like restriction we imposed on the local dynamics, 
$C^{1/2}(n,N)$ can at most be linear in $n$ as discussed in 
Ref.~\onlinecite{lt}.
This upper bound is reached by all the curves shown in the figure. Thus 
$C^{1/2}(n,N)\sim n^\chi $, with $\chi \approx 1$. However, the data
obviously contain additional dependence on the system size $N$ and suggest
the form $C^{1/2}(n,N)=n\cdot N^{\chi _{{\rm FS}}}$, where $\chi _{{\rm FS}}$
is the finite-size exponent. To obtain this exponent,

\begin{figure}
\narrowtext
\epsfxsize=3.0truein
\centerline{\epsffile{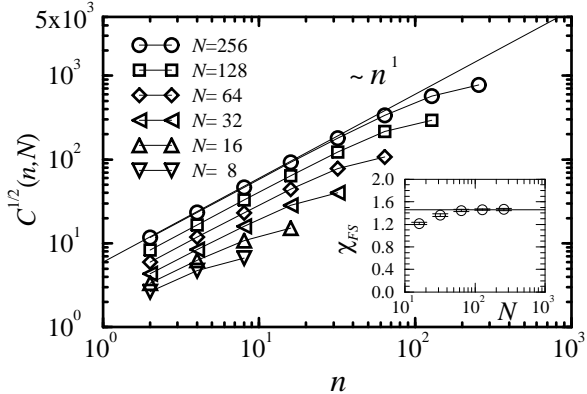}}
\vspace{\baselineskip}
\caption{
Correlation functions for the uniform chains,
plotted in the same way as Fig.~\ref{F15}.
}
\label{F16}
\end{figure}

\noindent  we compute an
effective exponent $\chi _{FS}(N)$ defined as~\cite{lt}
\begin{eqnarray}
\chi _{{\rm FS}}(N=2^i)=
\frac {1}{i-1}\sum_{j=2}^{i}
\log_2 \left( \frac{
C^{1/2}(2^j,N)}{C^{1/2}(2^{j-1},N/2)}\right).  \label{E3.8}
\end{eqnarray}
The result is shown in the inset of Fig.~\ref{F15}. 
We see $\chi _{{\rm FS}}(N)$ stabilize for $N>100$, 
yielding $\chi _{{\rm FS}}=1.22\pm 0.01$ for the random
chain. The same calculations finds $\chi _{{\rm FS}}=1.46\pm 0.01$ 
for the uniform spring chain (Fig.~\ref{F16}).

The obtained value for $\chi _{{\rm FS}}$ in the case of uniform chain is in
good agreement with the FRG result~\cite{nf}  which found
$\chi_{{\rm FS}} = (4-D + \eta_s)/2$, with $\eta_s = 0$ to one-loop order.
It is also  consistent  with the numerical result of  
$\chi _{{\rm FS}}=1.3\pm 0.3$ found in Ref.~\onlinecite{mf}.
The result for the random chain is also very close to the one reported in 
Ref.~\onlinecite{lt} for the simulation of a driven elastic string 
with random-field or
random-bond disorder ($\chi _{{\rm FS}}\simeq 1.25$), and to the ones
obtained from the simulations of related models of interface depinning 
($\chi _{{\rm FS}}=1.23\pm 0.02$ in Ref.~\onlinecite{R28}, 
and $\chi _{{\rm FS}}=1.2\pm 0.1$ in Ref.~\onlinecite{R29}). 
Combining the results on the two
independent exponents $\beta $ and $\chi $, we find strong evidence
supporting our expectation that the critical depinning dynamics of the
driven random chain is in the same universality class as the driven elastic
string in $(1+1)$-dimensional random medium. This is the $D=1$ case of our
more general conjecture concerning the critical dynamics of the $D$%
-dimensional randomly-tethered elastic network. Numerical studies of the
dynamics of the two-dimensional system is already underway~\cite{zeng}.

\section{Bulk-mediated Elasticity}

\begin{figure}
\narrowtext
\epsfxsize=2.0truein
\centerline{\epsffile{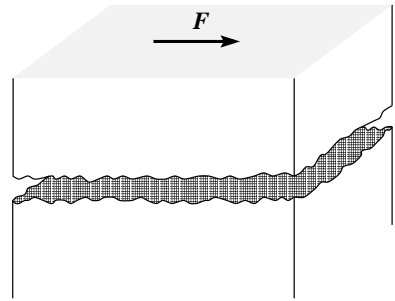}}
\vspace{\baselineskip}
\caption{
Picture illustrating two ``tectonic plates'' sliding past each other
along a ``fault zone".
}
\label{F17}
\end{figure}

So far we have studied the properties of the randomly-tethered elastic
network which is {\em completely} in contact with a disordered substrate. In
many situations however, the elastic medium
interacts with the substrate only at one of its surfaces. This is for
example the case of friction between a thick piece of rubber and a piece of
sandpaper. Similar models have been used to describe aspects of 
``tectonic plate'' movement along an earthquake fault zone 
(See Fig.~\ref{F17}). Here we extend the model of Sec.~IV to 
include bulk-mediated nonlocal
elasticity, and study the driven dynamics of such systems. 

\begin{figure}
\narrowtext
\epsfxsize=3.0truein
\centerline{\epsffile{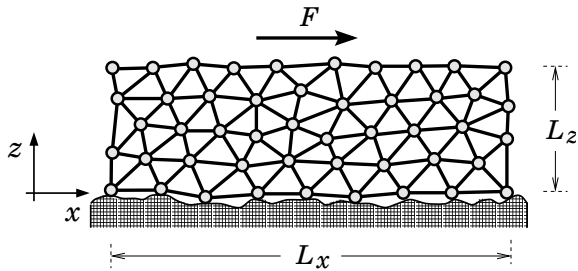}}
\vspace{\baselineskip}
\caption{
A 2D bead-spring system of thickness $L_z$ on 
disordered one-dimensional substrate.
}
\label{F18}
\end{figure}

We shall focus on the $(1+1)$-dimensional caricature of the problem depicted
in Fig.~\ref{F18}. We consider a two-dimensional random bead-spring network
(of size $L_x\times L_z$) in contact with a one-dimensional disordered (and
impenetrable) substrate $V(x)$ located at $z=0$ (see Fig.~\ref{F18}). We
wish to study the dynamics generated by a driving force applied to the upper
($z=L_z$) surface. For simplicity, we assume there is a sufficiently large
loading force $-F_N\hat{z}$ which keeps the elastic network in contact with
the substrate, thereby allowing us to suppress the $z$-degrees of freedom in
the displacement vectors $\bu$.

Let the $x$-component of $\bu({\bf R}_n)$ at the upper and lower boundaries 
of the
array be $w_0(n)$ and $w_1(n)$ respectively. Consider first the homogeneous
elastic medium. The equations of motion for the $w$'s are then 
\begin{eqnarray}
\mu^{-1}_0\partial _tw_0 &=&\sum_{n^{\prime }}\left\{ -G^{-1}(n-n^{\prime
},0)w_0(n^{\prime },t)\right.   \nonumber \\
&&\quad \quad \left. -G^{-1}(n-n^{\prime },L_z)w_1(n^{\prime },t)\right\}  
\nonumber \\
&&\quad -V'(n\cdot a+w_0(n,t))  \label{2spring.1} \\
\mu^{-1}_1\partial _tw_1 &=&\sum_{n^{\prime }}\left\{ -G^{-1}(n-n^{\prime
},0)w_1(n^{\prime },t)\right.   \nonumber \\
&&\quad \quad \left. -G^{-1}(n-n^{\prime },L_z)w_0(n^{\prime },t)\right\} +F
\label{2spring.2}
\end{eqnarray}
where $\mu _i$'s are the microscopic frictional coefficients, 
and the $G(n,z)$'s are the Green's functions whose
Fourier transforms are given in Appendix B. We assume that the upper
surface $w_1$ relaxes very quickly such that the nontrivial dynamics is
dominated by the behavior of the lower surface $w_0$, due to its contact
with the substrate. Setting $\partial _tw_1=0$ and solving 
Eq.~(\ref{2spring.2}), we obtain a closed equation of motion for $w_0$: 
\begin{eqnarray}
\mu _0\partial _tw_0 &=&\sum_{n^{\prime }}J(n-n^{\prime })w_0(n^{\prime },t)
\nonumber \\
&&\quad -V'(n\cdot a+w_0(n,t))+F,  \label{eom.f0}
\end{eqnarray}
where the kernel $J(n)$ is given in term of its Fourier transform 
\begin{eqnarray}
\widehat{J}(k) &=&-2\gamma L_z\,\left/ \sum_{n=-\infty }^\infty \frac 1{%
k^2+p_n^2}\right. \,,\quad p_n= n \pi {L_z}   \label{hJ.0} \\
&\approx &\left\{ 
\begin{tabular}{ll}
$-2\gamma \left| k\right| $ & ${\rm for}\quad \left| k\right| L_z\gg 1$ \\ 
$-2\gamma L_zk^2$ & ${\rm for}\quad \left| k\right| L_z\ll 1$%
\end{tabular}
\right. ,  \label{hJ.1}
\end{eqnarray}
as detailed in the Appendix B. Thus, the familiar form of bulk-mediated
elasticity, $\widehat{J}(k)\sim -|k|$ is recovered in the limit
of large $L_z,$ while for small $L_z$, the problem is effectively
one-dimensional in the limit of small $k$.

Randomness in the elastic medium can be readily incorporated into the
dynamics. Since the ground state configuration of the beads are {\em %
unfrustrated}, the effect of random spring lengths can be shifted away
completely upon using the displacement field $u({\bf R}_n)$ defined with
respect to the equilibrium position ${\bf R}_n=(x_n,z_n)$ of the beads in
the random system (without the external potential $V(x)$). Using 
$w_0(n)=u_0(n)+x_n-n\cdot a$ in Eq.~(\ref{eom.f0}), where 
$u_0(n)=u(x_n,z_n=0) $, we obtain finally the full equation of motion 
\begin{eqnarray}
\mu_0^{-1}\partial_t u_0&=&\sum_{n^{\prime }}J(n-n^{\prime })u_0(n',t) 
\nonumber \\
&&\quad -V'(x_n+u_0(n,t))+F.  \label{eom.f}
\end{eqnarray}

The coarse-grained dynamics generated by the discrete system (\ref{eom.f})
with the kernel (\ref{hJ.0}) can be derived as before. The resultant
equation of motion is 
\begin{eqnarray}
\mu^{-1}_0\,\partial_t u_0(x,t)&=&\int_{x'} \,J(x-x')u_0(x',t)\nonumber\\
& &\quad -V'(x)+f_{\rm pin}(x,u_0)+F,  \label{contact.eom}
\end{eqnarray}
where the pinning force is $f_{\rm pin}(x,u_0)
=V_>'(x)\delta \rho _0(x-u_0,z=0)$, with 
$\delta \rho _0(x,z)$ being density variation 
of the relaxed bead-spring system as
described before, yielding
\begin{equation}
\overline{[f_{\rm pin}(x,u_0)f_{\rm pin}(0,0)]} 
\approx \Delta \delta(x) \delta(u_0),
\label{contact.ff}
\end{equation}
for a variety of randomly-tethered networks.

Eq.~(\ref{contact.eom}) with the disorder correlator (\ref{contact.ff}) is
very similar to the equation of motion of a {\em contact line} which
controls the spreading of a non-wetting liquid on a disordered solid 
surface~\cite{dg}. As shown by Joanny and de Gennes~\cite{jdg}, 
a contact line is governed by
nonlocal elasticity of the form (\ref{hJ.1}), reflecting the energetics of
distorting the underlying liquid/gas interface. The driven dynamics of the
contact line has been investigated by Ertas and Kardar~\cite{ke3} 
using the FRG method. 
A depinning transition similar to that of the driven RM was
found. Upon generalizing the substrate to $D$-dimensions, an expansion about
the upper critical dimension $D=2$ yields the exponent values $\chi
=\epsilon /3$ and $z=1-2\epsilon /9$ to first order in $\epsilon =2-D$.
Other exponents can be obtained from the exponent relations 
$\nu =1/(1-\chi) $, $\beta =(z-\chi )/(1-\chi )$, 
and $\kappa =D-1+\chi $. Thus the contact
line problem ($\epsilon =1$) is described by the exponents $\nu \approx 3/2$, 
$\beta \approx 7/9$, and $\chi \approx 1/3$. Our problem of course differs
from that of the contact line again by the long-range correlations in higher
moments of the pinning force $f(x,u_0)$. However, as in the case of the
random spring with local elasticity, we do not expect these higher moments
to change the universality class of the depinning dynamics.

\begin{figure}
\narrowtext
\epsfxsize=3.0truein
\centerline{\epsffile{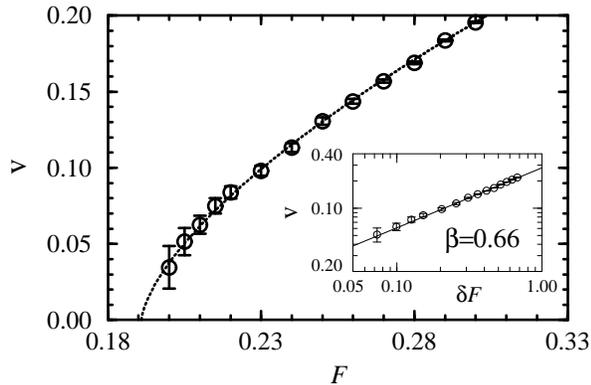}}
\vspace{\baselineskip}
\caption{
The velocity-force characteristics in the vicinity
of the depinning threshold, for the $1+1$ dimensional system 
of size $128\times 4096$. Log-log plot is shown in the inset.
The slope of
the straight line gives $\beta =0.66$.
}
\label{F19}
\end{figure}

\begin{figure}
\narrowtext
\epsfxsize=3.0truein
\centerline{\epsffile{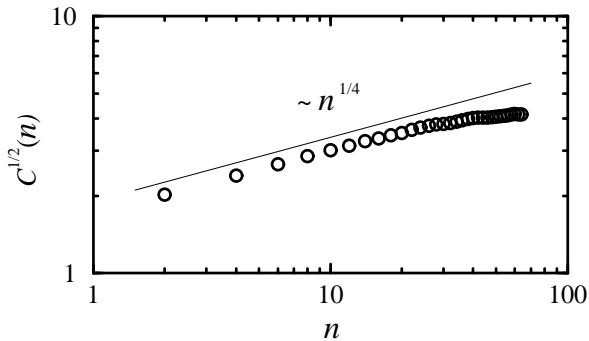}}
\vspace{\baselineskip}
\caption{
Roughness of the displacement field at the threshold $F_c=0.1910$. The
full line gives $\chi =0.25$. The statistical uncertainties are smaller than
symbol size.
}
\label{F20}
\end{figure}

To test this conjecture, we performed numerical integration of the discrete
equation of motion (\ref{eom.f}), with the effective kernel $\widehat{J}$
precomputed using (\ref{hJ.0}), with $L_x=128$ and $L_z=4096$. (Similar
results were obtained when we directly used the kernel 
$\widehat{J}= -2\gamma |k|$.) The
parameters describing the springs and the random potential $V$ were the same
as the ones used in Sec.~IV.B.  The data for the velocity-force
characteristic (shown in Fig.~\ref{F19}) were obtained by averaging over 
$15 $ independent samples, while the spatial correlation function of $u_0(n)$
at $F=F_c$ (Fig.~\ref{F20}) was averaged over $100$ samples. From the
numerical data, we find the exponents $\beta \approx 0.66$ and $\chi \approx
0.25$, which are consistent with the approximate FRG results for the contact
line. Thus the correspondence between the critical dynamics of the random
spring chain and that of the directed path appears to hold even with nonlocal
elasticity~\cite{N6}. 

We have so far discussed only the critical dynamics of the elastic medium
driven at a constant force. Another way the system may be driven is by a
constant velocity imposed at the upper boundary. Such situations are of
interest to the study of earthquakes~\cite{cls}, 
as they model tectonic plate motion
along a fault. The well-known Burridge-Knopoff model~\cite{bk} 
is of this class.

\begin{figure}
\narrowtext
\epsfxsize=2.6truein
\centerline{\epsffile{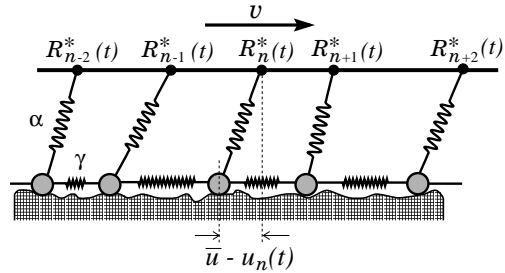}}
\vspace{\baselineskip}
\caption{
A one-dimensional random spring chain pulled by a constant velocity drive
via a set of loading springs of spring constant $\alpha$.
The mean lag distance is $\bar{u}$.
}
\label{F21}
\end{figure}

To consider the effect of a constant-$v$ drive, we return first to the linear
bead-spring chain described in Sec.~IV.B. We drive the chain by connecting
each bead $n$ to a reference point $R_{n}^{\ast}$ via a weak loading spring of
spring constant $\alpha $. The spacing of the reference points are chosen
such that when the beads are in the relaxed state (i.e., without the
external potential $V$), the loading springs are unstretched. Thus, 
$R_{n}^{\ast}=R_n=\sum_{m=1}^n a_m$ where $a_m$'s are the lengths of 
the springs connecting the beads. The reference points are then set to motion, with 
$R_{n}^{\ast}(t)=R_n+vt$. This motion forces the bead-spring chain to move, at
the {\em same} velocity $v$, via the action of the loading springs 
(see Fig.~\ref{F21}.)

The equation of motion for the beads is 
\begin{eqnarray*}
\partial _tr_n &=&\gamma [r_{n+1}-2r_n+r_{n-1}-(a_{n+1}-a_n)] \\
&&-\alpha [r_n-R_{n}^{\ast}(t)] -V'(r_n),
\end{eqnarray*}
where $r_n(t)$ is the position of the bead $n$. In term of the displacement
field $u_n(t)=r_n(t)-R_{n}^{\ast}(t)+\bar{u}$, ($\bar{u}$ is the mean {\em lag}
distance as shown in Fig.~\ref{F21}), the equation of motion becomes 
\begin{eqnarray}
\partial _tu_n &=&\gamma \left[ u_{n+1}-2u_n+u_{n-1}\right] -\alpha u_n 
\nonumber \\
&&\quad -v+\alpha \bar{u} -V'(R_n+vt+u_n(t)-\bar{u}).  \label{eom.v1}
\end{eqnarray}
It is instructive to compare Eq.~(\ref{eom.v1}) with Eq.~(\ref{E3.4})
evaluated in the moving frame $r_n(t)=R_n+vt+u_n(t)$: 
\begin{eqnarray}
\partial _tu_n &=&\gamma \left[ u_{n+1}-2u_n+u_{n-1}\right]  \nonumber \\
&&\quad -v -V'(R_n+vt+u_n(t))+F.  \label{eom.v0}
\end{eqnarray}
We find the systems described by (\ref{eom.v1}) and (\ref{eom.v0}) to be
statistically the same, up to the damping term $-\alpha u_n$, if we identify 
$\alpha \bar{u}$ in (\ref{eom.v1}) with the driving force $F$.
(The random forces $V'(x)$ and $V'(x-\bar{u})$ clearly have the same 
statistics.) The knowledge
that Eq.~(\ref{eom.v0}) exhibits a depinning transition with $v\propto
(F-F_c)^\beta $ and a diverging correlation length $\xi \sim (F-F_c)^{-\nu }$
immediately lead us to conclude that large fluctuations also occur in the
system with constant-$v$ drive as $v\rightarrow 0$, with 
$\xi \sim v^{-\nu/\beta }$ and $\bar{u} \to F_c/\alpha$. 
However, true critical behavior is prevented by the linear
damping term $-\alpha u_n$ in (\ref{eom.v1}). The correlation length
saturates at $\xi ^{*}\sim \left( \gamma /\alpha \right) ^{1/2}$.

We now turn to the system depicted in Fig.~\ref{F18}, with
the upper ($z=L_z $) boundary set to constant velocity.
Mathematically, the situation is
described by Eq.~(\ref{2spring.1}), with $w_1(n,t)=vt$. The explicit form of
the dynamics is readily obtained from the knowledge of $G^{-1}$ (see
Appendix B); we find the following equation of motion, 
\begin{eqnarray}
\partial _tu_0(n,t) &=&\sum_{n'} J(n-n')\,u_0(n',t)  \nonumber \\
&&\quad -\alpha _L(u_0-vt) -V'(x_n+u_0(n,t))  \label{eom.v2}
\end{eqnarray}
where the kernel $J(n)$ is approximately the same as that describing the
constant-$f$ dynamics in Eq.~(\ref{eom.f}), and the damping coefficient is
$\alpha _L\approx \gamma /(\pi L_z)$. In the moving frame 
$u_0\rightarrow u_0+vt-\bar{u}$, Eq.~(\ref{eom.v2}) 
has the same form as (\ref{eom.v1}) except for the nonlocal elasticity
coupling $J$. Thus, we expect similar ``near-critical'' behavior for the
system (\ref{eom.v2}) as $v\rightarrow 0$, with a diverging correlation
length $\xi \sim v^{-\nu /\beta }$ which saturates at $\xi ^{*}\approx
(\gamma /\alpha _L)^{1/\sigma }$. The value of the exponent $\sigma $
depends on which regime of $J$ the system is in at the scale $\xi ^{*}$;
see Eq.~(\ref{hJ.1}). We have $\sigma =2$ if $L_z/\xi ^{*}\ll 1$ 
and $\sigma =1$ if $L_z/\xi ^{*}\gg 1$.
Given the expression for $\alpha _L(L_z)$, we find $\sigma =1$. Thus, 
$\xi ^{*}\sim L_z$ becomes very large and the system is very nearly critical
as $v\rightarrow 0$. (Note that the lag distance $\bar{u}$ also
becomes very large.)
Zero-temperature driven dynamics with $v\to 0$ studied here is an example
of ``extremal dynamics'' studied extensively in the context of self-organized
critical phenomena~\cite{pmb}. 
The relation between self-organized criticality
and dynamic critical phenomena has been discussed
in various contexts~\cite{cbo,tb,nm,fisher,pb}.
Within our model spring-bead system, we see that the equivalence between 
these two classes of phenomena can be explicitly established.

We finish this section with a discussion of the ($D+1$)-dimensional
generalization of the system depicted in Fig.~\ref{F18}. Our analysis lead
us to anticipate strongly that such a system is equivalent 
to the $D$-component, 
$D$-dimensional driven manifold in $(D+D)$-dimensional random media, with a
nonlocal elasticity kernel (\ref{hJ.1}). In particular, the $D=2$ case is
analogous to the sliding slabs of elastic media (Fig.~\ref{F17}). Since 
$D=2 $ is the critical dimension for problems with bulk-mediated elasticity,
the critical exponents at the depinning transition are known 
exactly~\cite{ke1}, e.g., 
$\chi =O(\log )$, $\nu =1$, and $\beta =1$. From these and the exponent
relations given above,  one finds the exponent describing the avalanches 
at the onset of motion to be $\kappa =1$. 
Recently, the avalanche distribution of a related earthquake model was
investigated~\cite{fisher}. 
In the study of earthquakes, one monitors the distribution of
the ``moment'' $M\equiv \int d^D{\bf x\,}\Delta u({\bf x})$, where 
$\Delta u({\bf x})$ is the total displacement at location ${\bf x}$ during an
avalanche. Since $\chi =0$ in $D=2$, $M\sim \ell ^2$ where $\ell $ is the
linear size of the avalanche, assuming that the avalanche clusters are 
{\em compact} up to logarithmic correction. From (\ref{avalanche}), 
it follows~\cite{N7}
that the probability $\Pr (M)$ of finding an earthquake with moment
exceeding $M$ is $\Pr (M)\sim M^{-B}\widehat{\rho }(M/\xi ^2)$ at the
critical point, with $B=\kappa /2=1/2$. 
The model of Ref.~\onlinecite{fisher} uses constant-$v$
drive with $v\rightarrow 0$. Thus, $\xi =\xi ^{*}\sim L_z$.  The
numerical result on the moment-distribution is consistent with the RM
analogy, as has been pointed out in Ref.~\onlinecite{fisher}.

\section{Summary}

We have presented a detailed study of the static and dynamic properties
of an elastic medium interacting with a disordered substrate. The behavior
depends crucially on the equilibrium density distribution of the elastic medium
in the relaxed state. The interaction of a perfectly homogeneous medium 
with the substrate belongs to the CDW universality class as has already
been discussed in the literature. The somewhat surprising result of
this study is that a slight amount of {\em quenched-in} inhomogeneities
of the elastic medium, even with only interstitials/vacancies or 
phonons, is sufficient to change the universality class. Instead of the CDW,
a $D$-dimensional inhomogeneous medium on $D$-dimensional disordered
substrate belongs to the universality class of a $D$-dimensional
homogeneous manifold embedded 
in an effective $(D+D)$-dimensional random medium.
This is a consequence of the fact that quenched-in 
density variation breaks the translational symmetry of the elastic medium,
such that the dense/dilute parts of the medium preferentially stick to
the attractive/repulsive parts of the substrate.
We verified this numerically for a one-dimensional random
bead-spring system: The finite-temperature static behaviors of the 
random chain are found to be indistinguishable from those of the 
directed path in $(1+1)$-dimensional random medium. The zero-temperature
driven dynamics exhibits a depinning transition, the critical properties 
of which are also indistinguishable from those of the driven elastic 
string in $1+1$ dimensional
random medium. The equivalence is found to hold also for elastic media
with  nonlocal (bulk-mediated) elasticity, making  our model system
and results relevant to tribology problems. Finally, a slow-velocity drive
is shown to be equivalent to a constant-force drive close to the depinning
transition, demonstrating explicitly that self-organized 
critical phenomenon obtained from certain extremal dynamics may be viewed
as a dynamic critical phenomenon.

Although we formulated the interaction of the elastic medium with
the substrate in terms of the energetics of density variations, 
the underlying physics governing the interaction is much more general. 
The situation
being studied here is really one of {\em pattern matching}, i.e., 
matching of quenched-in fluctuations between two different elastic
media. This could occur as well in the form of roughness matching,
say, the matching of two rough surfaces in contact, or sequence matching,
as in the hybridization of two heterogeneous DNA sequences. By viewing  
interacting random systems  as  effective {\em homogeneous} systems
embedded in higher spatial dimension with external randomness,  it is
possible to simplify and resolve a large class of interesting problems. 
These include for example the reptation of heteropolymers in disordered
gel matrix~\cite{reptate}, 
where ``pattern matching'' of the  polymer composition 
with its  reptation tube leads to anomalously slow dynamics, and 
the preferential adsorption of heteropolymers on surfaces coated
with specific chemical patterns~\cite{adsorb}. 
We hope that this work will stimulate
further progress in understanding the physics of interacting random
systems.

\acknowledgements

We are grateful to P.~Bak, C.~Carraro, D.S.~Fisher, M.~L\"{a}ssig, 
 and D.R.~Nelson for helpful suggestions and comments.
T.H. acknowledges the hospitality of Ecole Normale Superieur where part
of this work was completed. This research is supported by an A.P. Sloan
research fellowship, and by an ONR
Young Investigator Award, through grant no. ONR-N00014-95-1-1002.

\vspace{24pt}
\noindent {\it Note added:} After the submission of this manuscript,
we became aware of recent works by K.V. Samokhin~\cite{samokhin}, 
who considered theoretically the case of an amorphous manifold 
(with local elastcity) on random substrate.
Conjecture on the perturbative irrelevancy of long-range correlated 
effective random potential was asserted based on a replica calculation.

\appendix
\section{Coarse-Grained Dynamics}

In this appendix we derive the  coarse grained  equation of motion
(\ref{eom.u}) (with the definitions (\ref{bbf}) -- (\ref{E2.9})),
starting from the  discrete model of the 
inhomogeneous elastic network driven by a constant force.
For simplicity, we shall describe only  the  case $D=2$, and furthermore
exclude all quenched-in topological defects including interstitials and
vacancies.
Inhomogeneities in the medium are described by small deviation
$\bw_{n_1,n_2}$ of the equilibrium positions of the beads 
from a perfect lattice, $\cR_{n_1,n_2} = n_1 \ba_1 + n_2 \ba_2$,
where the $\ba_i$'s are the lattice vectors. For $|\bw| \ll |\ba|$,
we can use $\bn\equiv(n_1,n_2)$ to label the beads.

We start with the {\em coarse-grained} density field,
\begin{equation}
\trho({\bf r},t)=\frac{1}{\Lambda^2} \int_\Lambda d^2 \br'
\, \rho(\br+\br')
\label{rho1.app}
\end{equation}
where 
\begin{equation}
\rho(\br) = \sum_{\bn} \delta\left(\br-{\bf r}_\bn(t)\right) \label{rho2.app}
\end{equation}
is the microscopic density field,
$\br_\bn(t)$ gives the actual position of the bead $\bn$, and $\Lambda$ 
is the coarse-graining scale, of the order of several $|\ba|$'s.
In Sec.~II, we showed that in terms of the displacement field $\bu(\br)$,
\begin{equation}
\rho(\br) \approx \bar{\rho} (1 - \grad \cdot \bu) 
+ \delta \rho_0(\br-\bu(\br)),
\label{rho3.app}
\end{equation}
where $\bar{\rho}=1/|\ba_1 \times \ba_2|$ is the average bead density,
and $\delta\rho_0(\bx)$ describes the equilibrium density variations.
For the specific 2D model considered here, 
\begin{equation}
\delta \rho_0(\bx) \sim \brho \sum_j \cos[\bK_j\cdot (\bx + \bw(\bx))],
\label{drho.app}
\end{equation}
with $\bK_j $ being the reciprocal lattice vectors.
Since $\delta\rho_0(\bx)$ fluctuates predominantly at the scale
$|\ba|$'s, it does not survive the coarse graining, and we have
\begin{equation}
\trho \approx \brho \, (1 - \grad \cdot \bu) \label{trho}
\end{equation}
to leading order in $(\grad \cdot \bu)$.

The equation of motion for $\bu$ can now
be obtained from the evolution  of $\rho$, which is given by the
continuity equation,
\begin{equation}
\partial_t\trho(\br,t) + \grad\cdot \tj(\br,t)=0,  \label{continuity}
\end{equation}
where $\tj$ is the coarse grained  ``current'', given by
\begin{equation}
\tj(\br,t)=\frac{1}{\Lambda^2} \int_\Lambda d^2 \br' \, \bj(\br+\br',t),
\label{j1.app}
\end{equation}
and
\begin{equation}
\bj(\br,t) = \sum_{\bn} \partial_t \br_{\bn} 
\, \delta\left(\br - \br_\bn(t)\right).\label{j2.app}
\end{equation}
This choice of the current automatically satisfies the continuity equation,
as can be verified directly by inserting the definitions 
(\ref{rho1.app})--(\ref{rho2.app}), and
(\ref{j1.app})--(\ref{j2.app}) into Eq.~(\ref{continuity}).
Using (\ref{trho}) for $\trho$ in Eq.~(\ref{continuity}), we easily find
the form of the equation of motion for $\bu$:
\begin{equation}
\partial_t \bu(\br,t) = \tj(\br,t)/\bar{\rho}. \label{eom.u0}
\end{equation}
Thus our task is to obtain the coarse-grained current $\tj$
starting from the discrete equation of motion,
\begin{eqnarray}
\mu_0^{-1} \partial_t \br_\bn 
&=& \sum_{\ba}
\gamma \big(\br_{\bn+\ba} - \br_\bn + \bw_{\bn+\ba} -\bw_\bn\big)
 + \bbf(\br_\bn), \label{eom.discrete}
\end{eqnarray}
where the sum is over the nearest neighbors, and
\begin{equation}
\bbf(\br) = -\grad V(\br) + {\bf F}
\label{def.f}
\end{equation}
is the external force.

Consider first the ``one-particle" equation of motion,
$ \mu_0^{-1} \partial_t \br_n = \bbf(\br_n(t))$.
Then from the definition (\ref{j1.app}), we have
\begin{eqnarray}
\bj(\br,t) &=& \mu_0 \rho(\br,t) \bbf(\br) \nonumber \\
&=& \mu_0 (1 - \grad \cdot \bu)\,  \bbf(\br) + \mu_0 \bbf_1(\br,\bu(t)), 
\label{bj.app}
\end{eqnarray}
where
\begin{equation}
\bbf_1(\br,\bu) = \delta\rho_0(\br-\bu) \cdot \bbf(\br)/\brho. 
\label{f1.app}
\end{equation}
Upon coarse graining of Eq.~(\ref{bj.app}), we find
\begin{equation}
\tj = \mu_0 (1 - \grad \cdot \bu) \, \tf(\br) + \mu_0\, \fpin(\br,\bu(t)), 
\label{tj.app}
\end{equation}
where 
\begin{equation}
\tf(\br) \equiv \Lambda^{-2} \int_\Lambda d^2\br' \,  \bbf(\br+\br')
= -\grad V_<(\br) + {\bf F}
\label{tf0.app}
\end{equation}
is the coarse-grained force, $V_<(\br)$ being the slowly varying part 
of the substrate potential, and
\begin{equation}
\fpin(\br,\bu) = -\grad V_>(\br) \, \delta\rho_0(\br-\bu),
\label{fpin.app}
\end{equation}
$V_>(\br)$ denoting the Fourier modes close to the reciprocal lattice
vector $\bK_i$'s.

Inclusion of bead-bead coupling as specified by Eq.~(\ref{eom.discrete}) 
only affect the term $\tf$.
For a statistically isotropic array (i.e., a triangular lattice), we find
\begin{eqnarray}
\tf &=& (c_{11}(\br)-c_{66}(\br)) \grad (\grad \cdot \bu) 
\nonumber \\
& & \quad + c_{66}(\br) \nabla^2 \bu 
-\grad V_<(\br) + {\bf F}
\label{tf.app}
\end{eqnarray}
where the elastic moduli $c_{11}, c_{66} \sim \gamma a^2$. (The
$\br$-dependences of the $c$'s come from $\grad\cdot\bw(\br)$ and
$\grad \times \bw(\br)$.)
The equation of motion for the displacement field $\bu$ is finally
\begin{equation}
\mu_0^{-1} \partial_t \bu = (1-\grad\cdot\bu)\, \tf + \fpin(\br,\bu)
\label{eom.u.app}
\end{equation}
with $\tf$ and $\fpin$ given by Eqs.~(\ref{tf.app}) and
(\ref{fpin.app}) respectively.

\section{Bulk-Mediated Elasticity}

In this appendix, we derive the nonlocal equation describing the motion of
the beads at the $z=0$ boundary of an elastic medium placed in $x-z$
half-plane (see Fig.~\ref{F18}). Our strategy is to obtain first the {\em %
effective} Hamiltonian describing the equilibrium fluctuation of the beads
at the $z=0$ boundary, subject to various drive conditions applied to the
opposing ($z=L$) boundary. The effective equation of motion is then obtained
by applying gradient descent dynamics.

We consider here the {\em homogeneous} elastic medium which is described by
the Hamiltonian (\ref{H0}). To simplify the description, we assume that a
sufficiently large normal force is applied such that displacement in the
transverse ($z$) direction is negligible. The Hamiltonian describing the
residual displacement fluctuation $w(x,z)$ in the $x$-direction can then be
written as 
\begin{equation}
\beta {\cal H}_0\left[ u\right] =\int dx\,dz\,\frac \gamma 2\left( %
\grad w(x,z)\right)^2.  \label{H0.app}
\end{equation}
The probability to find a system in the state with displacement $w_0(x)$ at
the $z=0$ boundary and $w_1(x)$ at the $z=L_z$ boundary is

\begin{eqnarray}
P\left[ w_0,w_1\right] &\propto &\int {\cal D}w\,\delta \left(
w(x,0)-w_0(x)\right)  \nonumber \\
&&\quad \quad \times \;\delta \left( w(x,L_z)-w_1(x)\right) \,e^{-\beta 
{\cal H}_0\left[ w\right] }.  \label{EB1}
\end{eqnarray}
Integrating out the $w$-field, we obtain

\begin{equation}
P\{w_0,w_1\}\propto e^{-\beta {\cal H}_{{\rm eff}}}  \label{EB3}
\end{equation}
where 
\begin{eqnarray}
\beta {\cal H}_{{\rm eff}} 
&=&\int \frac{dk}{2\pi }\left\{ \frac 12G^{-1}(k,0)\left[ \left| 
\widehat{w}_0(k)\right| ^2+\left| \widehat{w}_1(k)\right| ^2\right]
\right.   \nonumber \\
&&\qquad \qquad \quad \left. -G^{-1}(k,L_z)\widehat{w}_0(k)\widehat{w}%
_1(-k)\right\} ,  \label{EB4}
\end{eqnarray}
with $\widehat{w}_i(k)$ being the Fourier transform of $w_i(x)$, 
\begin{eqnarray}
G^{-1}(k,0) &=&\frac{C(k,0)}{C^2(k,0)-C^2(k,L_z)}  \label{EB5} \\
G^{-1}(k,L_z) &=&\frac{C(k,L_z)}{C^2(k,0)-C^2(k,L_z)}  \label{EB6}
\end{eqnarray}
and 
\begin{eqnarray}
C(k,z) &=&\frac 1{2\gamma L_z}\sum_{p=-\infty }^\infty 
\frac{\cos(pz)}{k^2+p^2}\,,\quad p=0,\pm \frac \pi {L_z},\ldots  
\nonumber \\
&\approx &\left\{ 
\begin{tabular}{ll}
$\frac 1{2\gamma |k|}e^{-|k| z}$ & ${\rm for}\quad |k| L_z\gg 1$ \\ 
$\frac 1{2\gamma L_zk^2}\left( 1+k^2L_z^2/\pi \right) $ & ${\rm for}%
\quad | k| L_z\ll 1$%
\end{tabular}
\right.   \label{C.app}
\end{eqnarray}
The equations of motion for $w_0$ and $w_1$ are now straightforwardly
obtained from gradient descent of $\beta {\cal H}_{{\rm eff}}$. For a
constant force drive applied to the $z=L_z$ boundary, we have 
\begin{eqnarray}
\mu^{-1}_0\partial _t\widehat{w}_0(k,t) &=&-G^{-1}(k,0)\widehat{w}_0(k,t) 
\nonumber \\
&&\quad -G^{-1}(k,L_z)\widehat{w}_1(k,t)  \label{EB8.a} \\
\mu^{-1}_1\partial _t\widehat{w}_1(k,t) &=&-G^{-1}(k,L_z)\widehat{w}_0(k,t)
\nonumber \\
&&\quad -G^{-1}(k,0)\widehat{w}_1(k,t)+F\,\delta (k)  \label{EB8.b}
\end{eqnarray}
where $\mu _i$ are the different microscopic frictional coefficients at the
boundaries. Assuming that the upper boundary $w_1$ relaxes very quickly such
that the nontrivial dynamics is dominated by the lower boundary (due to its
contact with the substrate), we can set $\partial _t w_1=0$ and obtain an
effective equation for $u_0$ itself. It reads in Fourier space 
\begin{equation}
\mu _0\partial _t\widehat{w}_0(k,t)=\widehat{J}(k)\widehat{w}%
_0(k,t)+F\,\delta (k)  \label{eom.f.app}
\end{equation}
with 
\begin{eqnarray}
\widehat{J}(k) &=&-C^{-1}(k,0)  \nonumber \\
&\approx &\left\{ 
\begin{tabular}{ll}
$-2\gamma \left| k\right| $ & ${\rm for}\quad \left| k\right| L_z\gg 1$
\\ 
$-2\gamma L_zk^2$ & ${\rm for}\quad \left| k\right| L_z\ll 1$%
\end{tabular}
\right. .  \label{hJ.app}
\end{eqnarray}
Thus, in the limit of large $L_z$, the kernel $\widehat{J}$ takes on the
form $\widehat{J}(k)=-2\gamma \left| k\right| $ familiar for
bulk-mediated elasticity. And in the opposite limit, the coupling becomes
local again as the system reverts back to being one-dimensional.

Equation of motion for constant $v$-drive at the $z=L_z$ boundary is
obtained simply by setting $\widehat{w}_1(k,t)=vt\,\delta (k)$ in Eq.~(%
\ref{EB8.a}). It reads in Fourier space 
\begin{eqnarray}
\partial _t\widehat{w}_0(k,t) &=&-G^{-1}(k,0)\widehat{w}_0(k,t) 
\nonumber \\
&&\quad +\;G^{-1}(k,L_z)v_0t\,\delta (k).  \label{EB8}
\end{eqnarray}
Using the result (\ref{C.app}) in Eqs.~(\ref{EB5}) and (\ref{EB6}), we find 
\begin{eqnarray}
G^{-1}(k,0) &=&\left\{ 
\begin{tabular}{ll}
$2\gamma \left| k\right| $ & ${\rm for}\quad \left| k\right| L_z\gg 1$
\\ 
$\alpha_L +\gamma _LL_zk^2$ & ${\rm for}\quad \left| k\right| L_z\ll 1$%
\end{tabular}
\right.  \label{G0.app} \\
G^{-1}(0,L_z) &=&\alpha_L  \label{G1.app}
\end{eqnarray}
with $\alpha_L \approx \gamma /(\pi L_z)$ and $\gamma _L\propto \gamma $. Thus
the equation of motion for the constant-$v$ drive is 
\begin{eqnarray}
\partial _t\widehat{w}_0(k,t) &\approx &-\alpha_L \left( \widehat{w}%
_0(k,t)-vt\delta (k)\right)  \nonumber \\
&&\quad +\widehat{J}(k)\widehat{w}_0(k,t),  \label{eom.v.app}
\end{eqnarray}
with $\widehat{J}(k)$ given approximately by Eq.~(\ref{hJ.app}).


\begin{references}

\vspace{-10pt}

\bibitem{ke1}See M. Kardar and D. Ertas, in {\it Scale Invariance, Interfaces,
and Non-Equilibrium Dynamics}, edited by A. McKane, M. Droz, J. Vannimenus,
and D. Wolf (Plenum Press, New York, 1995); and references therein.

\bibitem{cdw}H. Fukuyama and P.A. Lett, Phys. Rev. B {\bf 17},
535 (1978); P.A. Lee and T.M. Rice, Phys. Rev. B {\bf 19}, 3970 (1979).

\bibitem{cdw1}D.S. Fisher, Phys. Rev. B {\bf 31}, 1396 (1985).

\bibitem{nf}   O. Narayan
and D.S. Fisher, Phys. Rev. B {\bf 46}, 11520 (1992).

\bibitem{blatter}  G. Blatter, M. V. Feigel'man, V.B. Geshkenbein, 
A.I. Larkin, and V.M. Vinokur, Rev. Mod. Phys. {\bf 66}, 1125 (1994).

\bibitem{td}J. Toner and D.P. DiVicenzo, Phys. Rev. B {\bf 41}, 632 (1990).
\bibitem{ts} Y.-C. Tsai and Y. Shapir,  Phys. Rev. Lett. {\bf 69}, 1773 (1992);
Phys. Rev. E {\bf 50}, 3546 (1994); {\bf 50}, 4445 (1994).

\bibitem{front} R. Bruinsma and G. Aeppli, Phys. Rev. Lett. {\bf 52},
1547 (1984); J. Koplik and H. Levine, Phys. Rev. B {\bf 32}, 280 (1985).

\bibitem{co}  J.L. Cardy and S. Ostlund, Phys. Rev. B {\bf 25}, 6899 (1982).


\bibitem{psrg}  J. Villain and J.F. Fernandez, Z. Phys. B {\bf 54}, 139 (1984).

\bibitem{rft1} M. Mezard and G. Parisi, J. Phys. (Paris) I {\bf 1}, 809 (1991).

\bibitem{gld}T. Giamarchi and P. Le Doussal, 
Phys. Rev. B {\bf 52}, 1242 (1995).

\bibitem{frg}D.S. Fisher, Phys. Rev. Lett. {\bf 56}, 1964 (1986).

\bibitem{kor}S. Korshunov, Phys. Rev. B {\bf 48}, 3969 (1993).
\bibitem{tm}  M.~Kardar, and Y.-C.~Zhang, Phys. Rev. Lett. {\bf 58}, 2087
(1987).

\bibitem{dp}For a review of the directed path in random media, see
J. Krug and H. Spohn, in {\it Solids far from equilibrium: Growth,
Morphology and Defects}, C. Godreche ed. (Cambridge University Press, 1991);
and T.~Halpin-Healy, and T.-C.~Zhang, Phys. Rep. {\bf 254}, 215
(1995).

\bibitem{aam}A.A. Middleton, Phys. Rev. E. {\bf 52}, R3337 (1995).

\bibitem{mcmf}
C. Zeng {\it et al}, Phys. Rev. Lett. {\bf 77}, 3204 (1996);
H. Rieger and U. Blasum, Phys. Rev. B {\bf 55}, R7394 (1997).

\bibitem{natt}  T. Nattermann, S. Stepanow, L.-F. Tang, and H. Leschhorn, J.
Phys. II (France) {\bf 2}, 1483 (1992); H. Leschhorn, T. Nattermann, S.
Stepanow, L.-F. Tang, Annalen der Physik {\bf 6}, 1 (1997).

\bibitem{nf2}O. Narayan and D.S. Fisher, Phys. Rev. B {\bf 48}, 7030 (1993).

\bibitem{ke2} D. Erta\c{s} and M. Kardar, Phys. Rev. B {\bf 53}, 3520 (1996).

\bibitem{pbl}P. Sibani and P.B. Littlewood, Phys. Rev. Lett. {\bf 64},
1305 (1990).

\bibitem{mf}A.A. Middleton and D.S. Fisher, Phys. Rev. Lett. {\bf 66},
92 (1991), and Phys. Rev. B {\bf 47}, 3530 (1993).

\bibitem{myers}  C.R. Myers and J.P. Sethna, Phys. Rev. B {\bf 47}, 11171
(1993), {\it ibid.} {\bf 47}, 11194 (1993).

\bibitem{robbins}  B. Koiller, H. Ji, and M.O. Robbins, Phys. Rev. B {\bf 46},
5258 (1992); C.S. Nolle, B. Koiller, N. Martys, and M.O. Robbins, Phys. Rev.
Lett. {\bf 71}, 2074 (1993).

\bibitem{string}  M. Dong, M.C. Marchetti, A.A. Middleton, and V. Vinokur, Phys.
Rev. Lett. {\bf 70}, 662 (1993).

\bibitem{lt}  H. Leschhorn and L.-H. Tang, Phys. Rev. Lett. {\bf 70}, 2973
(1993); H. Leschhorn, Physica A {\bf 195}, 324 (1993).

\bibitem{stanley}L.A.N. Amaral, A.-L. Barabasi, and H.E. Stanley,
Phys. Rev. Lett. {\bf 73}, 62 (1994).

\bibitem{cule} D. Cule, Phys. Rev. E {\bf 52}, 1 (1995).

\bibitem{rubio}  M.A. Rubio, C.A. Edwards, A. Dougherty, and J.P. Gollub, Phys.
Rev. Lett. {\bf 63}, 1685 (1989).

\bibitem{hor} V.K. Horv\'{a}th, F. Family, and T.
Vicsek, Phys. Rev. Lett. {\bf 67}, 3207 (1991).

\bibitem{wong} S. He, G.L.M.K.S. Kahanda,
and P.-Z. Wong, Phys. Rev. Lett. {\bf 69}, 3731 (1992).

\bibitem{bu} S.V. Buldyrev {\it et al}, Phys. Rev. A {\bf 45}, R8313 (1992).

\bibitem{family}F. Family, K.C.B. Chan, and J. Amar, in {\it Surface 
Disordering: Growth, Roughening and Phase Transitions}, Les Houches Series,
(Nova, New York, 1992).

\bibitem{bk}  R. Burridge and L. Knopoff, Bull. Seismol. Soc. Am. {\bf 57},
341 (1967).

\bibitem{taka}H. Takayasu and M. Matsuzaki, Phys. Lett. A {\bf 131}, 
244 (1988).

\bibitem{cl}J. Carlson and J.S. Langer, Phys. Rev. Lett. {\bf 62}, 2632 (1989);
 Phys. Rev. A {\bf 40}, 6470 (1989).

\bibitem{naka}H. Nakanishi, Phys. Rev. A {\bf 41}, 7086 (1990).


\bibitem{lub}P.A. Thompson and M.O. Robbins, Science {\bf 250}, 
792 (1990).

\bibitem{feder}  H.J.S. Feder, and J. Feder, Phys. Rev. Lett. {\bf 66}, 2669
(1991).

\bibitem{gollub}  D.P. Vallette, and J.P. Golub, Phys. Rev. E {\bf 47}, 820
(1993).

\bibitem{cls}  J.M. Carlson, J.S. Langer, and B.E. Shaw, Rev. Mod. Phys. 
{\bf 66}, 657 (1994).


\bibitem{bt}P. Bak and C. Tang, Geophys. Rev. B {\bf 94}, 15635.

\bibitem{cbo}K. Chen, P. Bak, and S.P. Obukhov, Phys. Rev. A {\bf 43},
625 (1991).

\bibitem{tb}C. Tang and P. Bak, Phys. Rev. Lett. {\bf 60}, 2347 (1988).

\bibitem{btw}  P. Bak, C. Tang, and K. Wiesenfeld, Phys. Rev. Lett. {\bf 59}%
, 381 (1987).

\bibitem{twbcl}C. Tang, K. Wiesenfeld, P. Bak, S. Coppersmith, 
and P. Littlewood,  Phys. Rev. Lett. {\bf 58}, 1161 (1987).

\bibitem{nm}O. Narayan and A.A. Middleton, Phys. Rev. B {\bf 49}, 244 (1994).

\bibitem{sornette}P. Miltenberger, D. Sornette and C. Vanette, Phys. Rev. Lett.
{\bf 71}, 3604 (1993); P.A. Cowie, C. Vanette and D. Sornette, J. Geophys.
Res. {\bf 98}, 21809 (1993). 



\bibitem{prl}  D.~Cule, and T.~Hwa, Phys. Rev. Lett. {\bf 77}, 278 (1996).

\bibitem{fisher}D.S. Fisher, K. Dahmen, S. Ramanathan and Y. Ben-Zion,
Phys. Rev. Lett. {\bf 78}, 4885  (1997).

\bibitem{pb}M. Paczuski and S. Boettcher, 
Phys. Rev. Lett. {\bf 77}, 111 (1996).

\bibitem{hmv} T. Hwa, M.C. Marchetti, and V.M. Vinokur (unpublished).

\bibitem{reptate}D. Cule and T. Hwa, preprint (cond-mat/9706262).

\bibitem{align}T. Hwa and M. L\"{a}ssig, Phys. Rev. Lett. 
{\bf 76}, 2591 (1996).

\bibitem{pmb}M. Paczuski, S. Maslov, and P. Bak, Phys. Rev. E {\bf 53},
414 (1996).


\bibitem{landau}L.D. Landau and E.M. Lifshitz, {\it Theory of Elasticity},
(Pergamon Press, London, 1959).


\bibitem{N2}The elastic moduli have space-dependent components 
due to the presence 
of quenched randomness in the medium; however short-ranged variations
are irrelevant as can be easily checked.

\bibitem{N3}A random torque coupling to $\grad \times \bu$
will be generated by the interaction and may be included in Eq.~(\ref{H.eff}) 
for completeness. However, this and the random compression term 
$\bar{\rho}\,(\grad\cdot\bu) V(\br)$ are both
irrelevant as we will see.

\bibitem{cn}  C.~Carraro, and D.R.~Nelson, Phys. Rev. E {\bf 56}, 797 (1997).

\bibitem{cld}D. Carpentier and P. LeDoussal, Phys. Rev. B {\bf 55}, 12128
(1997).

\bibitem{rsb}P. Le Doussal and T. Giamarchi, Phys. Rev. Lett. {\bf 74},
606 (1995); J. Kierfeld, J. Phys. (France) I {\bf 5}, 379 (1995).

\bibitem{vgnum}G.G. Batrouni and T. Hwa, Phys. Rev. Lett. {\bf 72}, 
4133 (1994); D. Cule and Y. Shapir, Phys. Rev. Lett. {\bf 74}, 114 (1995)
and Phys. Rev. B {\bf 51}, 3305 (1995); H. Rieger, Phys. Rev. Lett. {\bf 74},
4964 (1995); E. Marinari, R. Monasson, J.J. Ruiz-Lorenzo, J. Phys. A {\bf 28},
3975 (1995).


\bibitem{hf}T. Hwa and D.S. Fisher, Phys. Rev. Lett. {\bf 72}, 2466 (1994).


\bibitem{thh}T. Halpin-Healy, Phys. Rev. Lett. {\bf 62}, 442 (1989);
Phys. Rev. A {\bf 42}, 711 (1990).

\bibitem{lb}For a recent review, see L. Balents, 
Lecture notes for the Beg-Rohu Spring School on Disordered Systems,
available online at http://www.itp.ucsb.edu/~balents/bignotes/bignotes.html.


\bibitem{N4} For the random manifold, linear
terms such as $\brho \, (\grad\cdot\bu) V({\bf r})$ 
in the Hamiltonian (\ref{H.eff}) are irrelevant.

%\bibitem{collapse}If we further include the 
%excluded-volume effect of the beads in the manifold, then such collapse
%will not be possible at all.

\bibitem{nv}D.R. Nelson and V.M. Vinokur, Phys. Rev. B {\bf 48}, 13060 (1993).

\bibitem{hnv}T. Hwa, D.R. Nelson, and V.M. Vinokur, Phys. Rev. B {\bf 67}, 
1167 (1993).


\bibitem{khh}J. Krug and T. Halpin-Healy, J. Phys. I (France) {\bf 3}, 
2179 (1993); I. Arsenin, J. Krug and T. Halpin-Healy, Phys. Rev. E
{\bf 49}, R3561 (1994).


\bibitem{hn}N. Hatano and D.R. Nelson, Phys. Rev. Lett. {\bf 77}, 570
(1996).

\bibitem{ml}In Ref.~\onlinecite{align}, the irrelevancy of correlated
disorder for a closely related sequence alignment problem was conjectured
based on similar reasons.

\bibitem{N1}  More generally, if the dimensionality of ${\bf u}$ and $\br$ 
are $D_u$ and $D_r$ respectively, then the problem is equivalent to that
of a $D_u$-component random manifold in $D_r$-dimensions.

\bibitem{zeng}C. Zeng {\it et al}, private communication (1997).

\bibitem{drn}We are grateful to D.R. Nelson for suggestions and discussions.
See also Ref.~\cite{cn} for discussions.

\bibitem{cut}C. Carraro and D.S. Fisher, Phys. Rev. B {\bf 51}, 534 (1995).


\bibitem{dsf} We are indebted to D.S. Fisher for an illuminating discussion
on the significance of the relabeling symmetry breaking in the context of the 
one-dimensional random-spring system.

\bibitem{gld1} A similar disorder correlator was obtained in Appendix A
of Ref.~\cite{gld}.


\bibitem{tm1}  M. Kardar, in {\it Fluctuating Geometries in Statistical
Mechanics and Field Theory - Les Houches}, Vol. 62, edited by F. David, P.
Ginsparg, and J. Zinn-Justin (Elsevier, Amsterdam, 1996).

\bibitem{gi}E. Medina, T. Hwa, M. Kardar and Y.-C. Zhang, 
Phys. Rev. A {\bf 39} 3053, (1987).

\bibitem{fh}D.S. Fisher and D.A. Huse, Phys. Rev. B {\bf 43} 10728, (1991).

\bibitem{kmb}  J.M.~Kim, M.A.~Moore, and A.J. Bray, Phys. Rev. A {\bf 44},
2345 (1991); J.M.~Kim, A.J. Bray, and M.A. Moore, Phys. Rev. A {\bf 44},
R4782 (1991).

\bibitem{exp}The avalanche distribution was discussed 
in the context of CDW in Ref.~\onlinecite{nm}, and extended
to RM in Ref.~\onlinecite{nf2} . The exponent relation for
$\kappa$ relies on argument relating the critical behaviors as
$F\to F_c^+$ and $F \to F_c^-$.

\bibitem{2drm}H.~Ji and M.0.~Robbins,  Phys. Rev. B {\bf 46}, 14519 (1992).

\bibitem{R19}  J.~Krug, Phys. Rev. Lett. {\bf 75}, 1795 (1995).

\bibitem{R20}  L.~Balents and M.P.A.~Fisher, Phys. Rev. Lett. {\bf 75}, 4270
(1995).

\bibitem{gld2} Some of the terms have also been obtained in T.
Giamarchi and P. Le Doussal, Phys. Rev. Lett. {\bf 76}, 3408 (1996).

\bibitem{gradv} The $\grad V$ term is used  in RG calculations~\cite{cld} as
a ``bookkeeping'' device to keep track of fluctuations in ${\bf u}$.


\bibitem{aniso} For certain anisotropic driven system~\cite{stanley}, 
the coefficient
of the $(\grad\cdot\bu)$ terms does not vanish even as $\tf \to 0$; 
see Ref.~\onlinecite{R10} for a detailed discussion. This is however
not the case here.

\bibitem{R10}  L.-H Tang, M. Kardar, and D. Dhar, Phys. Rev. Lett. {\bf 74},
920 (1995).

\bibitem{N5} The kinetic term 
${\bf f} (\grad \cdot \bu)$ is again irrelevant right at the
depinning transition as in the CDW case.


\bibitem{overlap}The regime of overlapping avalanches was studied in 
Ref.~\onlinecite{hk} for the sandpile model. The dynamics in this regime
was shown to be described by the driven-diffusion equation, which is 
a nonlinear Langevin equation. For the problem at hand, the moving phase
with overlapping avalanches is described analogously by the KPZ 
equation~\cite{kpz}.

\bibitem{hk}T. Hwa and M. Kardar, Phys. Rev. A {\bf 45}, 7002 (1992).

\bibitem{kpz}M. Kardar, G. Parisi, and Y.-C. Zhang, Phys. Rev. Lett. {\bf 56}, 
889 (1986).

\bibitem{R28}  S. Maslov, M. Paczuski, and P. Bak, Phys. Rev. Lett. {\bf 73}%
, 2162 (1994).

\bibitem{R29}  S. Roux and A. Hansen, J. de Physique I {\bf 4}, 515 (1994).

\bibitem{dg}  P.G. de Gennes, Rev. Mod. Phys. {\bf 57}, 827 (1985).

\bibitem{jdg}J.F. Joanny and P.G. de Gennes, J. Chem. Phys. {\bf 81},
552 (1984).


\bibitem{ke3}  D. Ertas, and M. Kardar, Phys. Rev. E {\bf 49}, R2532 (1994).


\bibitem{N6} To test our results more quantitatively, it would be
useful to compare directly with the numerics of the contact line
depinning when it becomes available.

\bibitem{N7} More
generally, $M\sim \ell ^{D_f+\chi }$, where $D_f\leq D$ is the fractal
dimension of the avalanche cluster, yielding $B=\kappa /(D_f+\chi )$.

\bibitem{adsorb}T. Hwa and D. Cule, preprint (cond-mat/9707079).

\bibitem{samokhin}K.V. Samokhin, JETP Lett. {\bf 64}, 580 (1996);
{\it ibid.} 853 (1996).



\end{references}
\end{document}